\documentclass[%
amsfonts, amsmath, amssymb,
aps, 
prx, 
twocolumn, 
10pt, 
preprintnumbers, 
floats, 
altaffillsymbol, 
final, 
reprint, 
balancelastpage, 
flushbottom, 
noraggedfooter, 
showpacs, 
nobibnotes, 
nofootinbib, 
longbibliography, 
superscriptaddress,
]{revtex4-2}
\usepackage{Preamble}
\usepackage{caption}
\usepackage{soul}
\usepackage{ragged2e} 

\usepackage{float}
\usepackage{graphicx}
\usepackage{mathrsfs}  
\usepackage{amsmath}
\usepackage{bm}
\usepackage{physics}
\usepackage{xcolor}
\definecolor{BLUE}{rgb}{0,0,1}

\usepackage{tikz}
\makeatletter
\renewcommand{\@makecaption}[2]{%
  \vskip\abovecaptionskip
  \begin{minipage}{\linewidth} 
  \small
  \parbox{\linewidth}{\justify #1. #2} 
  \end{minipage}
  \vskip\belowcaptionskip
}
\makeatother
\usepackage{placeins}
\usepackage{wrapfig}
\usepackage{hyperref}
\newcommand{\sect}[1]{\par\textit{#1}---}

\begin{document}

\title{Quasinormal modes of rotating black holes with ultralight scalar fields}

\author{Leif Lui}
\affiliation{Beijing Institute of Mathematical Sciences and Applications, Beijing 101408, China}
\author{Adrian Ka-Wai Chung}
\affiliation{St John's College, Cambridge, CB2 1TP, United Kingdom}
\affiliation{DAMTP, Centre for Mathematical Sciences, University of Cambridge, Wilberforce Road, Cambridge CB3 0WA, United Kingdom}
\email[Corresponding author: ]{kwc43@cam.ac.uk}

\date{\today}

\begin{abstract}
We report the first accurate computation of the leading-order shifts in the quasinormal-mode frequencies of rapidly rotating black holes surrounded by massive scalar fields with nonminimal curvature coupling. 
For fields with reduced Compton wavelengths no shorter than the black-hole mass scale, the numerical errors in the frequency shifts are typically $10^{-8} - 10^{-5}$ for dimensionless spins between 0 and 0.9, and at most $\mathcal{O}(10^{-3})$ near the scalar-mass threshold. Our results open up the possibility of searching for ultralight scalar particles, prime dark-matter candidates, using existing gravitational-wave signals.
\end{abstract}

\maketitle

\sect{Introduction}The nature of dark matter remains a central open question in fundamental physics. Astrophysical and cosmological observations infer its presence~\cite{Sofue:2000jx, Bertone:2016nfn, SupernovaCosmologyProject:1998vns, SupernovaSearchTeam:1998fmf}, yet its particle nature remains unknown despite extensive experimental searches. Ultralight bosons are a well-motivated class of dark-matter candidates~\cite{Weinberg_1977, Wilczek_1978, Hui_2017, Ferreira_2021}, with examples such as the QCD axion~\cite{Peccei_1977} that also address the strong-CP problem~\cite{Preskill_1982, Kim_2010}. Their macroscopic Compton wavelengths make black holes natural probes of such particles, motivating extensive gravitational-wave searches for dark matter~\cite{Arvanitaki_2011, Arvanitaki_2015}.

Most black-hole searches for ultralight bosons have focused on superradiance~\cite{Detweiler_1980, Dolan_2007, Siemonsen_2020, Baryakhtar_2021, Ng_2021, Tong_2022, Fan_2024, May_2025}, in which a bosonic cloud grows by extracting rotational energy from a black hole and produces observable gravitational-wave signatures~\cite{Brito_2020}. A distinct possibility arises when light scalar fields couple nonminimally to spacetime curvature. In this case, the black-hole curvature itself sources a nonvanishing scalar configuration, without relying on superradiant growth~\cite{Herdeiro_2018, Doneva_2018, Doneva_2024, Andreou_2019, Cunha_2019, Dima_2020, Belkhadria_2025}. Such theories can also be regarded as massive extensions of scalar-mediated theories beyond General Relativity~\cite{Berti_2015, Barack_2018, Yunes_2024}, which is natural given that no fundamental massless scalar particle is known in the Standard Model \cite{Aitchison:2003tq}. Motivated by these possibilities, numerous studies have attempted to probe ultralight dark matter through extreme mass ratio inspirals~\cite{Maselli_2021, Duque_2024, Wade_2026, Rahman_2026} via scalar radiation~\cite{Barsanti_2022, Zhang_2023, Dyson_2025, Li_2025, Barsanti_2026, Keijzer_2026}, gravitational atom spectroscopy~\cite{Cole_2023, Tomaselli_2023, Tomaselli_2024, Rocca_2026, Xu_2026a, Xu_2026b}, and by examining the transition-to-plunge orbits~\cite{Lui_2026c}. Despite this motivation, while the gravitational perturbations and quasinormal-mode spectra of rapidly rotating black holes have been studied in the context of massive minimally-coupled ~\cite{Berti_2009, Duque_2024, Spieksma_2025, Fonseca_2026, Berti_2026} and massless non-minimally-coupled scalar fields~\cite{Wagle_2022, Elley_2022, Chung_2024, Chung_2026a, Husken_2026}, the massive non-minimal case~\cite{Macedo_2018, Chatzifotis_2022, Richards_2023, D_Addario_2024, Lutfioglu_2025, Xie_2025, Chung_2026b} remains largely unexplored.

In this Letter, we compute the leading-order shifts in the quasinormal-mode frequencies of rotating black holes with massive curvature-coupled scalar fields, with numerical uncertainties well below those of current and projected ground-based ringdown measurements.
The shifts are obtained using a newly developed scheme that avoids the explicit symbolic derivation of the linearized field equations, enabling efficient calculations across the parameter space.
For dynamical Chern--Simons and scalar Gauss--Bonnet gravity, we obtain converged spectra for dimensionless spins up to $0.9$ and reduced Compton wavelengths comparable to or longer than the black-hole mass scale.
The frequency shifts exhibit distinct dependences on the scalar mass and on the parity of the curvature coupling, while we find no indication of instability among the modes explored. Our results establish the ringdown spectra needed to search for ultralight curvature-coupled scalar particles in gravitational-wave signals and extend our understanding of gravitational perturbations of rotating black holes carrying scalar matter.

 Throughout this Letter, we use geometrized units in which $c=G=\hbar=M=1$, where $M$ denotes the black-hole mass, modified Boyer-Lindquist coordinates $x^{\alpha} = (t, r, \chi = \cos \theta, \phi)$, and metric signature is $(-,+,+,+)$. In these units, $M_{\odot}{^{-1}}\simeq1.3\times10^{-10}\,\mathrm{eV}$.
\sect{Models of ultralight scalar particles}We consider the massive scalar particles governed by the following Lagrangian density~\cite{Cano_2019, Chung_2026a}
\begin{equation}\label{eq:Lagrangian}
\begin{split}
16 \pi \mathscr{L} = R & + \ell_1{}^2 \vartheta_1 \mathscr{G} - \frac{1}{2} \nabla_{\nu} \vartheta_1 \nabla^{\nu} \vartheta_1  - \frac{1}{2} \mu_1 {}^2 \vartheta_1 {}^2  \\
& + \ell_2{}^2 \vartheta_2 \mathscr{P} - \frac{1}{2} \nabla_{\nu} \vartheta_2 \nabla^{\nu} \vartheta_2 - \frac{1}{2} \mu_2 {}^2 \vartheta_2 {}^2, \\
\end{split}
\end{equation}
where $R$ is the Ricci scalar, and $\vartheta_{1,2}$ denote scalar and pseudoscalar fields with masses $\mu_{1,2}$ and coupling scales $\ell_{1,2}$.
We define the Gauss--Bonnet and Pontryagin invariants as 
$
\mathscr{G}=R^2-4R_{\mu\nu}R^{\mu\nu}
+R_{\mu\nu\rho\sigma}R^{\mu\nu\rho\sigma},$
and
$
\mathscr{P}
=
\tilde{R}^{\mu\nu\rho\sigma}
R_{\mu\nu\rho\sigma},
$
respectively, where $R_{\mu\nu}$, $R_{\mu\nu\rho\sigma}$, and
$\tilde{R}_{\mu\nu\rho\sigma}$ are the Ricci, Riemann, and dual Riemann tensors, respectively.
The limits $(\ell_1,\ell_2)=(\ell,0)$ and $(0,\ell)$ yield scalar Gauss-Bonnet and dynamical Chern-Simons couplings, respectively.
The case of $(\ell_1,\ell_2)=(\ell,\ell)$ corresponds to axi-dilaton coupling, which might be a possible mechanism to drive the cosmological inflation~\cite{Kallosh:2022vha}.

Varying the action gives the equations of motion,
\begin{equation}
R_\alpha{}^\beta
+\zeta\left(\mathscr{A}_\alpha{}^\beta-\bar T_\alpha{}^\beta\right)=0,
\qquad
\left(\Box-\mu^2\right)\bar{\vartheta}
+\mathscr{Q}=0,
\label{eq:field_equations}
\end{equation}
where $\zeta=(\ell/M)^4$ is the dimensionless coupling parameter, $\bar{\vartheta}=\ell^{-2}\vartheta$ is the rescaled scalar field, $\mathscr{Q}=\mathscr{P}$ for dynamical Chern-Simons coupling and $\mathscr{Q}=\mathscr{G}$ for scalar Gauss-Bonnet coupling, and $\bar T_\alpha{}^\beta$ is the trace-reversed energy-momentum tensor
of the rescaled scalar field.
The explicit forms of $\mathscr{A}_\alpha{}^\beta$ are given in the Supplemental Material.

We solve these equations for stationary and axisymmetric backgrounds using the metric ansatz of Ref.~\cite{Cano:2019ore}, together with the corresponding gauge choice and boundary conditions. The finite scalar mass introduces multiple radial scales in the background scalar configuration. We therefore construct the backgrounds using spectral collocation with radial domain decomposition. The numerical accuracy of the backgrounds and their contribution to the subsequent quasinormal-mode calculation are quantified in the Supplemental Material. We obtain rotating black-hole spacetimes and scalar configurations over a range of dimensionless spins and scalar masses in both dynamical Chern-Simons and scalar Gauss-Bonnet gravity. The resulting numerical backgrounds are publicly available at~\cite{Lui_Chung_2026}.

\sect{Black-hole perturbations}We perturb both the background spacetime and its surrounding massive
scalar field 
\begin{equation}\label{eq:perturbations}
\begin{split}
g_{\alpha\beta} = g_{\alpha\beta}^{(0)} +\varepsilon h_{\alpha\beta},
\qquad
\vartheta = \vartheta^{(0)} +\varepsilon\vartheta^{(1)},
\end{split}
\end{equation}
where $|\varepsilon|\ll1$ is a bookkeeping parameter, and we have suppressed the scalar-field label.
To compute the quasinormal-mode frequencies, we impose the Regge--Wheeler-type gauge used in Refs.~\cite{Chung_2023,Chung_2024,Chung_2024b,Lam_2026}, and its applicability in the presence of the massive curvature-coupled scalar field is discussed in the Supplemental Material.
In this gauge, the metric perturbation is described by six functions and is written as~\cite{Chung_2023,Chung_2024,Chung_2024b,Lam_2026}
\begin{equation}
h_{\alpha\beta}
=
\sum_{k=1}^{6}
h_k(r,\chi)\,
\tau^{(k)}_{\alpha\beta}(r,\chi)\,
\mathcal{P}_0(r)\,
e^{im\phi-i\omega t},
\label{eq:pert}
\end{equation}
where $\tau^{(k)}_{\alpha\beta}$ are template tensors, and $\mathcal{P}_0$ imposes the ingoing and outgoing quasinormal-mode boundary conditions at the horizon and infinity, respectively.
The functions $h_k(r,\chi)$ are bounded throughout the computational domain.

Similarly, the scalar perturbation can be written as
\begin{equation}\label{eq:Ps}
\vartheta^{(1)} = \frac{h_7(r,\chi)}{r}\, \mathcal{P}_s(r)\, e^{im\phi-i\omega t},
\end{equation}
where $h_7(r,\chi)$ is bounded and $\mathcal{P}_s$ incorporates the
massive-scalar asymptotic behavior. Defining the massive-scalar wave number $k=\sqrt{\omega^2-\mu^2}$, the free scalar solutions have asymptotic behavior $\sim e^{\pm i k r}$.
For the retarded branch, the scalar is outgoing in the open channel, $\mathfrak{Re}(\omega^{(0)})>\mu$, whereas the free scalar component becomes evanescent in the closed channel, $\mathfrak{Re}(\omega^{(0)})<\mu$.

\sect{Eigenvalue perturbation}The finite scalar mass introduces distinct asymptotic behaviours for
the gravitational and scalar perturbations through $k=\sqrt{\omega^2-\mu^2}$.
We solve the coupled linearized field equations using spectral collocation, with the corresponding boundary behaviour incorporated into the perturbation ansatz.
Details of the asymptotic factors, threshold treatment, and numerical implementation are detailed in the Supplemental Material. Here we summarize the eigenvalue-perturbation scheme~\cite{Hussain_2022, Green_2023}.

The spectral expansions of the bounded $h_{k}(r, \chi)$ can be schematically expressed as 
\begin{equation}
h_k=\sum_{n=0}^{N_z}\sum_{\ell}v_{kn\ell}\,\varphi_{n \ell}(z, \chi),
\end{equation}
where $v_{kn\ell}$ are the spectral coefficients, $\varphi_{n \ell}$ is a spectral basis of both $r$ and $\chi$, and $z$ is a compactified radial coordinate.
By defining $\textbf{v}=\{v_{kn\ell}\}$ and discretizing the linearized field equations, the quasinormal-mode problem becomes the eigenvalue problem \cite{Chung:2024vaf}
\begin{equation}
\mathbf{D}(\omega,\zeta)\mathbf{v}=0,
\qquad
\mathbf{D}(\omega,\zeta)
=
\mathbf{D}^{(0)}(\omega)
+\zeta\,\mathbf{D}^{(1)}(\omega),
\label{eq:eigenproblem}
\end{equation}
where $\mathbf{D}^{(0)}$ is the zeroth-order operator evaluated using background spacetime in general relativity, and $\mathbf{D}^{(1)}$ contains the leading correction from the curvature coupling.
Unlike previous implementations of spectral methods~\cite{Chung_2023, Chung_2024, Chung:2024vaf, Chung_2024b, Chung:2025gyg}, the matrix coefficients are assembled numerically from the field equations at the collocation points, without first deriving their explicit symbolic linearization.

Expanding the eigenfrequency and eigenvector as $\omega=\omega^{(0)}+\zeta\omega^{(1)}$ and $\mathbf{v}=\mathbf{v}^{(0)}+\zeta\ \mathbf{v}^{(1)}$ gives \cite{Chung:2024vaf}
\begin{equation}
\mathbf{D}^{(0)}\left(\omega^{(0)}\right)\mathbf{v}^{(1)} + \omega^{(1)} \partial_\omega\mathbf{D}^{(0)}\left(\omega^{(0)}\right)\mathbf{v}^{(0)}
=
-\mathbf{D}^{(1)}\left(\omega^{(0)}\right)\mathbf{v}^{(0)}.
\label{eq:first_order_shift}
\end{equation}
We solve Eq.~\eqref{eq:first_order_shift} as a bordered least-squares problem, yielding the leading frequency shift $\omega^{(1)}$ from a single linear solve.
Further details of the operator construction and numerical implementation are given in the Supplemental Material.

\begin{figure*}[htp!]
    \centering
    \includegraphics[width=\linewidth]{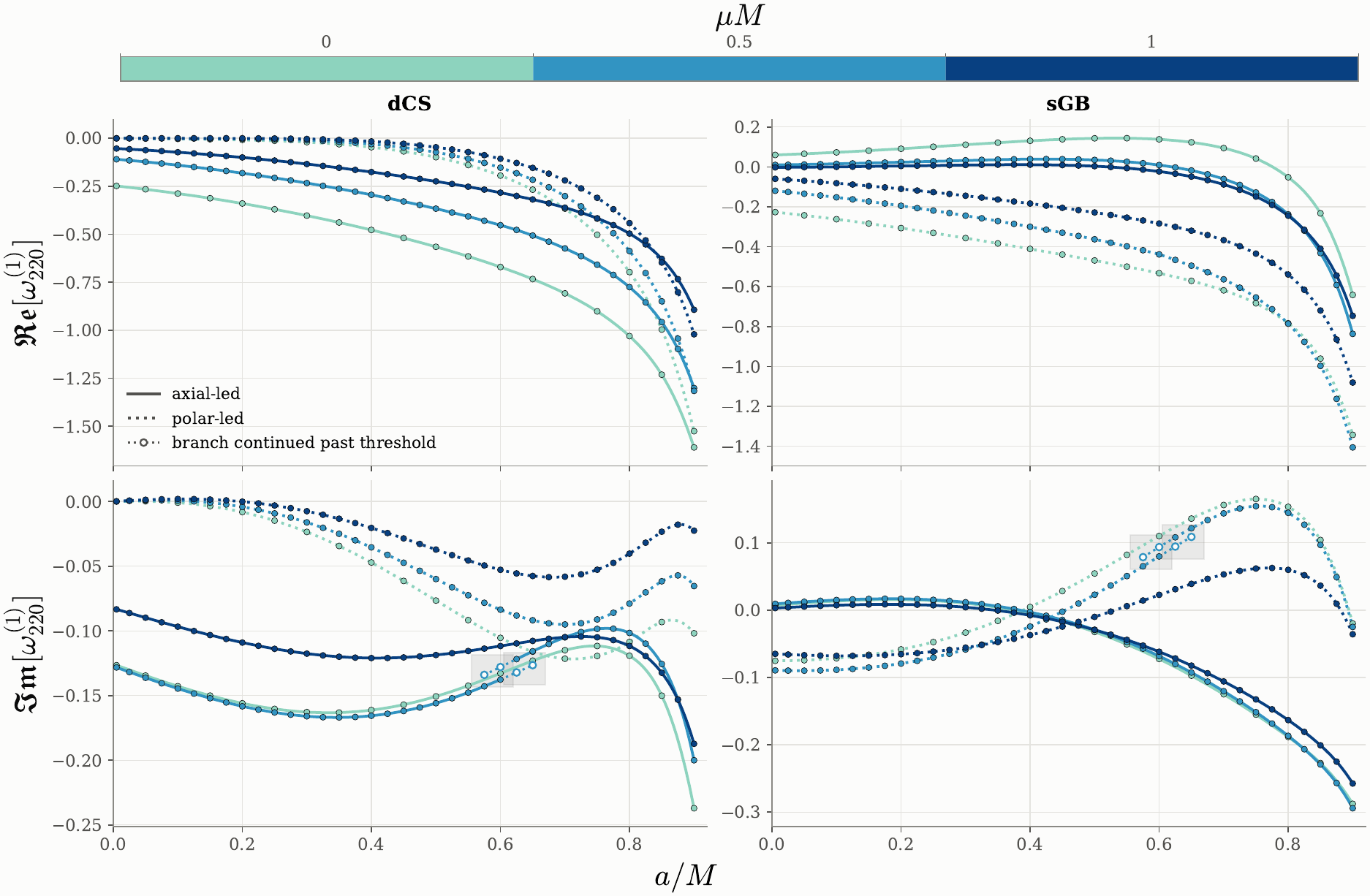}
    \caption{Leading-order shift $\omega^{(1)}_{220}$ of the frequency of the fundamental (2,2,0) quasinormal mode against the dimensionless spin $a/M$, for dynamical Chern-Simons (left) and scalar Gauss-Bonnet (right) gravity. The top and bottom panels show the real and imaginary parts of $\omega^{(1)}_{220}$, respectively, in units of $M^{-1}$. The data, denoted by the markers, are computed values using METRICS. The colours give the scalar mass, $\mu M=0$, $0.5$ and $1$, and the solid and dotted lines are the cubic-spline interpolants for the axial- and polar-led modes, respectively. The hollow dots within the shaded boxes are the branch-continued parts of the quasinormal modes (c.f. Fig.~\ref{fig:threshold} in the Supplementary Material).}
    \label{fig:omega_1}
\end{figure*}

\sect{Results}We apply our method to compute the leading-order frequency shifts of the $(2,2,0)$, $(2,2,1)$, and $(3,3,0)$ modes.
The $(2,2,0)$ mode is the fundamental co-rotating quadrupolar mode and is typically dominant in the ringdown phase of binary black-hole signals, while the $(2,2,1)$ overtone has recently been identified in GW250114~\cite{LIGOScientific:2025wao}.
Figure~\ref{fig:omega_1} shows the real (top) and imaginary (bottom) parts of $\omega^{(1)}_{220}$ in dynamical Chern-Simons (left) and scalar Gauss-Bonnet (right) gravity. Results for the $(2,2,1)$ and $(3,3,0)$ modes are presented in the Supplemental Material.
Numerical uncertainties, typically $\lesssim10^{-5}$ (median $2\times10^{-6}$) and at most $2\times10^{-3}$ next to the threshold at high spin (Supplemental Material, Fig.~S2), are included as error bars in Fig.~\ref{fig:omega_1}, but are too small to be visually resolved on the scale of the figure.

We observe some features from Fig.~\ref{fig:omega_1}.
First, the frequency shifts depend strongly on parity, demonstrating the breaking of axial--polar isospectrality by the curvature coupling. As $a\rightarrow0$, the polar-led shift vanishes in dynamical Chern-Simons gravity because the background is then Schwarzschild, and polar perturbations decouple from the pseudoscalar. In scalar Gauss-Bonnet gravity the scalar hair is already present at $a=0$, so both parities shift. These features are also present in the massless limit~\cite{Chung_2024b,Chung:2024vaf,Chung:2025gyg}.
Second, for fixed parity, the spin dependence of both the real and imaginary parts remains qualitatively similar across different scalar masses, while their magnitudes vary substantially with $\mu$. Third, the spectra recover the massless results continuously as $\mu\rightarrow0$, with $\omega^{(1)}(\mu)-\omega^{(1)}(0)\propto\mu$ in scalar Gauss-Bonnet and $\propto\mu^2$ in dynamical Chern-Simons gravity, reflecting the monopole and dipole scalar hair. At $\mu=0$ our dynamical Chern-Simons shifts agree with Ref.~\cite{Chung_2025} to better than $1\%$ (Supplemental Material, Figs.~S3--S5).

The step in $\omega^{(1)}$ where $\mathfrak{Re}[\omega^{(0)}(a)]=\mu$ (the $\mu M=0.5$ curves of Fig.~\ref{fig:omega_1}, at $a/M\simeq0.617$) is a threshold transition with a simple origin. 
In the far-field regime, the scalar perturbation behaves as $\sim e^{-i\omega t+ikr}$ with $k^2=\omega^2-\mu^2$, so $k(\omega)$ has branch points at $\omega=\pm\mu$~\cite{Koyama_2001, Hod_1998, Burko_2004, Konoplya_2007}. By defining $\kappa\equiv(\omega_{\ell mn}^{(0)2}-\mu^2)^{1/2}$ with $\mathfrak{Re}(\kappa)>0$, a damped mode would always yield a root where $\mathfrak{Im}(\kappa)<0$. The retarded Green's function~\cite{Leaver_1986, Andersson_1997, Ching_1995} selects $k=\kappa$ (outgoing scalar radiation) for $\mathfrak{Re}(\omega^{(0)})>\mu$, and $k=-\kappa$ ($\mathfrak{Im}(k)>0$, an evanescent, bound scalar) for $\mathfrak{Re}[\omega^{(0)}]<\mu$. The first-order shift depends on the channel through this boundary condition, $\omega^{(1)}=\Omega(\pm\kappa)$. Splitting $\Omega$ into even and odd parts under $k\to-k$, gives
\begin{equation}\label{eq:threshold}
\begin{split}
\omega^{(1)}(a,\mu)&=A(a,\mu)+s\cdot\kappa\,B(a,\mu),
\end{split}
\end{equation}
where $s=\mathrm{sgn}\big[\mathfrak{Re}(\omega^{(0)})-\mu\big]$, $A,B$ are analytic through the threshold, because $\omega^{(0)}$ stays a distance $|\mathfrak{Im}(\omega^{(0)})|$ from the branch point. The only non-analyticity is the sign $s$, so $\omega^{(1)}$ is piecewise analytic with a jump
\begin{equation}\label{eq:jump}
\begin{split}
\Delta\omega^{(1)}=2\kappa_{\rm th}B,\quad\kappa_{\rm th}&=\left[2i\mu\,\mathfrak{Im}\left(\omega^{(0)}\right)-\mathfrak{Im}\left(\omega^{(0)}\right)^2\right]^{1/2}\\
&\simeq\left(2\mu\left|\mathfrak{Im}\left(\omega^{(0)}\right)\right|\right)^{1/2}e^{-i\pi/4},\\
\end{split}
\end{equation}
\begin{figure}
    \centering
    \includegraphics[width=\linewidth]{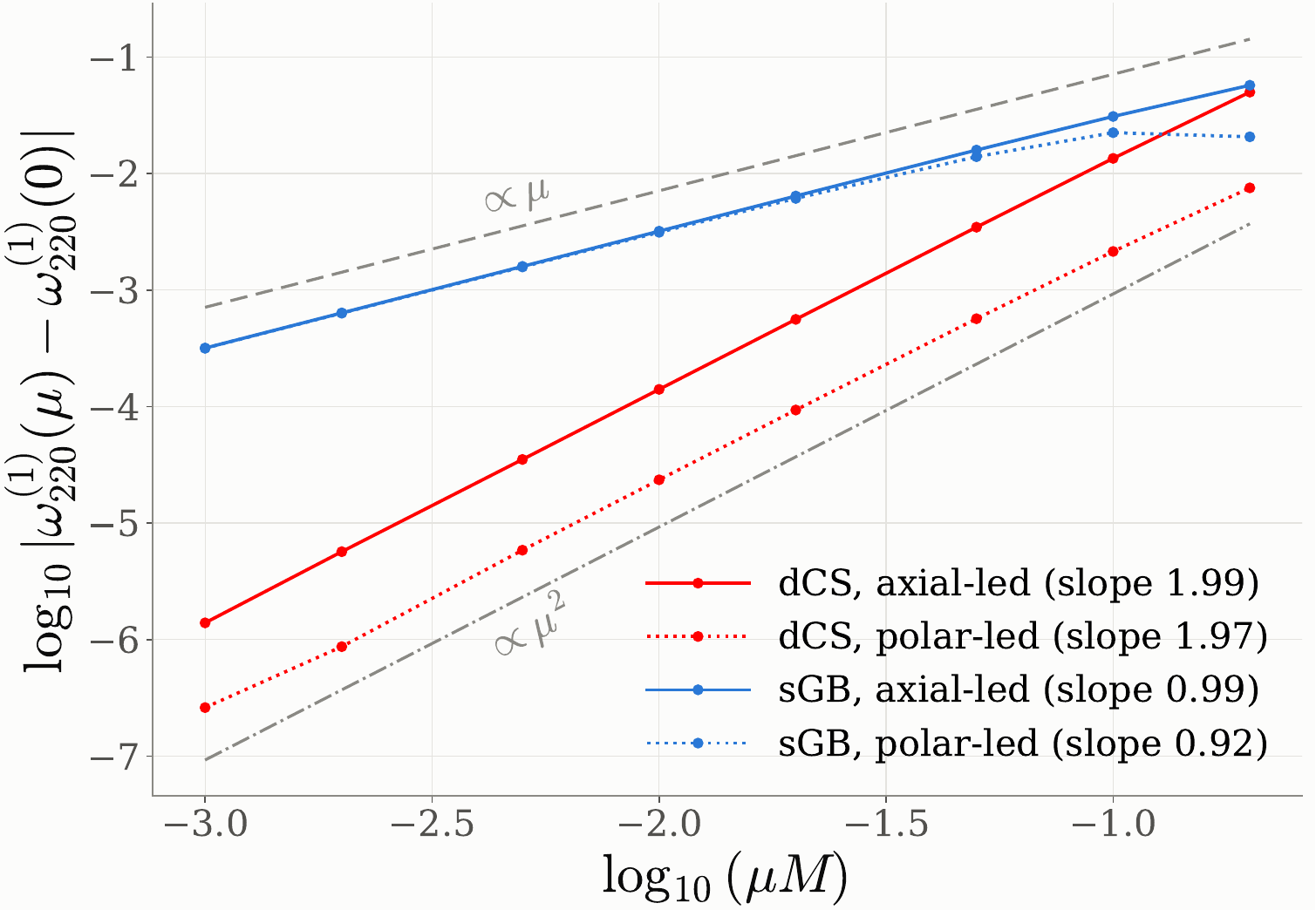}
    \caption{Approach to the massless limit at $a/M = 0.5$. $|\omega^{(1)}_{220}(\mu) - \omega^{(1)}_{220}(0)|$ against $\mu M$ for dynamical Chern-Simons (dCS, orange) and scalar Gauss-Bonnet (sGB, blue), axial-led (solid) and polar-led (dotted) modes.
Grey lines show $\propto\mu^2$ and $\propto\mu$, and the legend gives each curve's fitted slopes.}
    \label{fig:massless_limit}
\end{figure}
evaluated where $\mathfrak{Re}(\omega^{(0)})=\mu$. The jump is finite because the gravitational mode is damped, as $\kappa$ does not vanish at threshold ($|\kappa_{\rm th}|=0.289$ for $\mu M=0.5$). Its size is set by the odd, radiative part $B$ of the scalar response, which is large only in the parity that couples to the scalar at zero spin, axial in dynamical Chern-Simons and polar in scalar Gauss-Bonnet. At $\mu M=0.5$ we find $\Delta\omega^{(1)}_{220}=0.0008+0.0101\,i$ (dynamical Chern-Simons, axial-led) and $-0.0038+0.0138\,i$ (scalar Gauss-Bonnet, polar-led), against $\lesssim3\times10^{-5}$ in the other parity. In both cases, the jump is almost entirely in the damping rate. In the language of phase transitions, $\mathfrak{Re}(\omega^{(0)})-\mu$ is the control parameter and $s$ is the order parameter. The transition is of first order, and each branch continues smoothly past it, like a metastable phase. Each branch is a smooth curve in spin, and the physical mode hops from one to the other at threshold (c.f. Fig.~\ref{fig:threshold} in the Supplementary Material). The continued branches (hollow markers in Fig.~\ref{fig:omega_1}) are computed by imposing the other boundary condition. At finite coupling the transition sits where the pole itself crosses the cut~\cite{Ching_1995, Ohashi_2004, Konoplya_2005, Konoplya_2006, De_Amicis_2024, De_Amicis_2025}, $\mathfrak{Re}[\omega^{(0)}_{\ell mn}+\zeta\omega^{(1)}_{\ell mn}]=\mu$, which moves it to higher spin ($a/M\simeq0.77$ at $\zeta=0.1$ in both theories), while the ringdown signal itself stays continuous (c.f. Supplementary Material).
\begin{figure}[ht!]
    \centering
    \includegraphics[width=\linewidth]{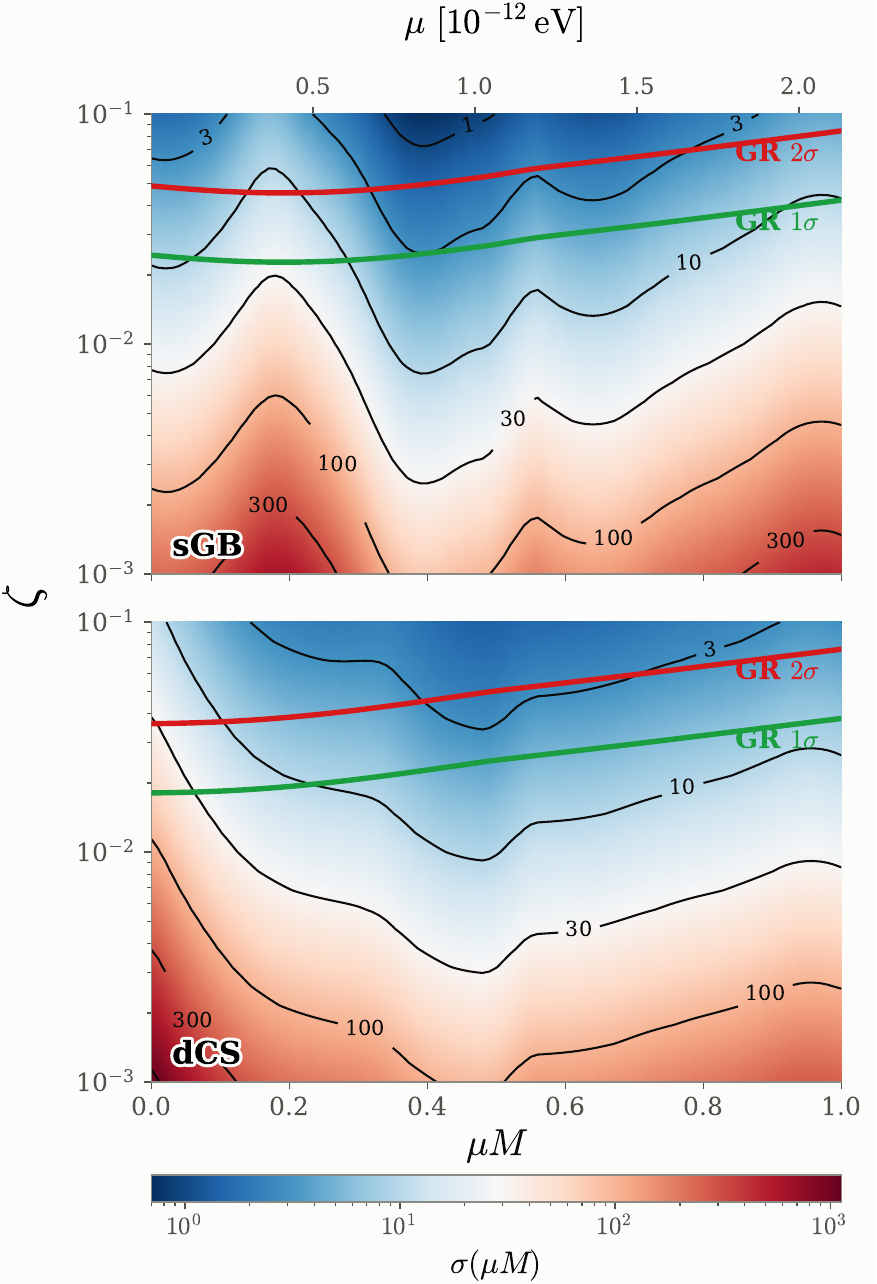}
    \caption{Forecast $1\sigma$ uncertainty on the scalar mass, $\sigma(\mu M)$, from the
ringdown of GW250114, as a function of the true scalar mass $\mu M$ and coupling $\zeta$,
for scalar Gauss-Bonnet (top; polar-led modes) and dynamical Chern-Simons (bottom; axial-led modes). We assume LVK O4
sensitivity, a ringdown signal-to-noise ratio $\rho = 26$ in the $(2,2,0)$ and $(2,2,1)$
tones, and a remnant spin $a/M = 0.68$. The remnant mass and spin are marginalised with
inspiral priors ($2\%$ in $M$, $0.02$ in $a/M$), as are the amplitudes and phases of both
tones. The green and red lines are the $1\sigma$ and $2\sigma$ ($68\%$ and $95\%$) upper
limits on $\zeta$ expected if the signal is consistent with General Relativity. The top axis gives $\mu$
for the source-frame remnant mass. Near $\mu M \approx 0.5$, where $\mu$ crosses
$\mathfrak{Re}(\omega^{(0)})$ of the two tones and the scalar channel closes, the map and
the limits are interpolated.}
    \label{fig:Fisher}
\end{figure}

\sect{Massless limit}
The results in Fig.~\ref{fig:omega_1} suggest that $\omega^{(1)}$ approaches the corresponding massless result as $\mu\rightarrow0$.
To examine this limit explicitly, we compute $\omega^{(1)}_{220}$ in both theories down to $\mu M=10^{-3}$.
Figure~\ref{fig:massless_limit} shows $\log_{10}|\omega^{(1)}_{220}(\mu)-\omega^{(1)}_{220}(0)|$ as a function of $\log_{10}(\mu M)$ for a black hole with $a/M=0.5$.
With both couplings, the difference decreases consistently as $\mu$ is reduced, demonstrating a continuous recovery of the massless limit.
Curiously, however, the two theories approach this limit with different scalings, as the fitted slopes are $0.92$ and $0.99$ in scalar Gauss-Bonnet but $1.97$ and $1.99$ in dynamical Chern-Simons gravity. That is, $\omega^{(1)}(\mu)-\omega^{(1)}(0)\propto\mu$ and $\propto\mu^2$, respectively.

This difference can be understood from the far-field structure of the massless scalar hair.
In scalar Gauss-Bonnet gravity, the leading scalar field is monopolar, $\bar\vartheta\simeq Q/r$, with charge $Q=4\sqrt{1-a^2}/(1+\sqrt{1-a^2})$~\cite{Yunes_2011, Prabhu_2018}. 
The finite mass of the scalar field screens this long-range profile at $r\sim\mu^{-1}$. 
Near the black hole the field only shifts by a constant, $\bar\vartheta\simeq Q/r-Q\mu$, which the shift-symmetric Gauss--Bonnet coupling does not feel\footnote{Strictly speaking, there are still additional terms $\sim \delta_{\alpha}{}^{\beta} \mu^2 \bar{\vartheta}^2$ in the trace-reversed energy-momentum tensor in the equations of motion, but their effects are at least of $\mathcal{O}(\mu^2)$.}, so the only $\mathcal{O}(\mu)$ effect is the field energy removed from the far zone, $\zeta Q^2\mu/8$~\cite{Sotiriou_2014}. At fixed ADM mass and angular momentum, the reduction in scalar-field energy is compensated by a shift in the black-hole mass $M$ and $a=J/M^2$ in the Kerr frequency, giving, for both parities,
\begin{equation}\label{eq:scalar Gauss-Bonnet_small_mu}
\omega^{(1)}(\mu)-\omega^{(1)}(0)=-\frac{Q^2}{8}\left(\omega^{(0)}+2a\,\frac{\partial\omega^{(0)}}{\partial a}\right)\mu+\mathcal{O}(\mu^2).
\end{equation}
This agrees with our scalar Gauss-Bonnet result to better than $0.6\%$ at $\mu M=10^{-3}$ for $a/M=0.1$--$0.5$, with a residual linear in $\mu$. This is expected from the $\mathcal{O}(\mu^2)$ term, as the nonanalytic linear term is a pure renormalisation of the black-hole mass.

With dynamical Chern-Simons coupling, the leading scalar hair is instead dipolar, $\bar\vartheta\sim r^{-2}$, for which the corresponding far-zone contribution scales as $\mu^3$ and therefore does not generate a linear term. 
The leading correction is then the ordinary $O(\mu^2)$ contribution associated with the scalar mass term and the massive scalar operator.

\sect{Implications for gravitational-wave searches for ultralight scalar particles}The spectra computed here make it possible to search black-hole ringdowns for ultralight scalars with nonminimal curvature couplings~\cite{Yamada_2019, Cano_2023, Silva_2023}.
To assess this, we perform a Fisher-information-matrix~\cite{Vallisneri_2008, Gair_2022, Wang_2022b, Dupletsa_2025} analysis of a GW250114-like ringdown observed by the LIGO--Virgo--KAGRA detectors at O4 sensitivity, modelled by the $(2,2,0)$ and $(2,2,1)$ tones (Fig.~\ref{fig:Fisher}).
We evaluate the noise-weighted inner product using the projected O4 power-spectral density \cite{Abbott_2020, Capote:2024rmo}.
The measurement uncertainty, $\sigma$, is estimated from the square root of the corresponding diagonal element of the inverse of the Fisher information matrix. At $2\sigma$, $\zeta\lesssim0.049$ (scalar Gauss-Bonnet) and $\zeta\lesssim0.036$ (dynamical Chern-Simons) as $\mu\to0$, weakening to $0.085$ and $0.076$ at $\mu M=1$ (i.e.\ $\mu\simeq2\times10^{-12}\,$eV for the $62.7\, M_\odot$ remnant). Meaning, if the signal is consistent with General Relativity ($\zeta=0$), the scalar mass cannot be measured, and the ringdown instead bounds the coupling for each assumed $\mu$. For couplings near these bounds, the scalar mass would be measured only to $\sigma(\mu M)\approx2$--$13$ ($\sigma(\mu)\approx4\times10^{-12}$--$3\times10^{-11}\,\rm eV$) for this remnant, wider than the range $\mu M\le1$ itself, so a single event of this loudness bounds $\zeta$ but cannot determine $\mu$. Since $\sigma(\mu)\sim1/\rho$, the scalar-mass uncertainty would decrease for louder signals or with next-generation detectors.

Despite the encouraging prospects suggested by this forecast, the Fisher-matrix analysis should be interpreted as indicative rather than
as a substitute for a full Bayesian parameter-estimation study.
The Fisher approximation assumes that the likelihood is locally Gaussian in the parameters and is asymptotically justified at high
signal-to-noise ratio.
At the ringdown signal-to-noise ratio of GW250114, departures from this approximation may be non-negligible.
Moreover, the physical boundary $\zeta\geq0$ violates the regular interior-parameter assumptions of the Fisher formalism, while at $\zeta=0$ the scalar mass is not identifiable and $\zeta$ and $\mu$ become strongly degenerate.
The perturbative frequency shift also changes branch where the scalar channel opens, so derivatives with respect to $\mu$ are not regular in the immediate vicinity of $\mu M\simeq0.5$.
Our forecast further assumes a fixed ringdown start time, a two-tone model with inspiral-informed priors on the remnant mass and spin, and a single representative detector noise curve.

\sect{Concluding remarks}In this work, we present the first accurate calculation of the leading-order shifts in the quasinormal-mode frequencies of rotating black holes surrounded by massive scalar fields with nonminimal curvature couplings.
Within the parameter space explored, we find no indication of an unstable quasinormal mode.
These results provide a basis for studying both the perturbative dynamics of such black holes and their gravitational-wave signatures in the ringdown phase of binary-black-hole mergers.

Our forecast shows that a single GW250114-like ringdown could constrain the coupling to $\zeta\lesssim0.04$--$0.09$ at $2\sigma$ for scalar masses $\mu\lesssim{}2\times10^{-12}\,\mathrm{eV}$, whereas measuring the scalar mass itself will require louder signals or next-generation detectors. 
These results open a new avenue for probing ultralight scalar particles directly through black-hole ringdown observations. To fully realize the potential of these searches, a full Bayesian parameter-estimation analysis of detector data, jointly sampling $\zeta$ and $\mu$ together with the remnant parameters using the spectra presented here, will be required to turn these forecasts into robust constraints. A dedicated search for massive ultralight scalar fields in detected gravitational-wave signals will be presented separately. 
Our numerical results also provide the ingredients for modelling extreme-mass-ratio inspirals around black holes carrying massive scalar hair, extending sensitivity to complementary scalar-mass ranges.

Beyond their observational implications, our numerical methods represent an advance in the computation of black-hole perturbations.
In contrast to our previous implementations of METRICS~\cite{Chung_2023,Chung_2024,Chung_2024b,Lam_2026}, which relied on explicit symbolic derivation of the linearized field equations, the present scheme assembles the perturbation operator numerically, substantially reducing the computational overhead associated with
extending the calculation across theories and parameter space.
For the production settings used here, a single frequency-shift calculation typically requires only $\mathcal{O}(5$--$10)$ minutes on a single CPU core, allowing spectra containing hundreds of parameter
points to be generated within hours using modest parallelization. 
Beyond this computational acceleration, our formalism consistently treats the transition between propagating and evanescent scalar channels, enabling calculations across a broad range of scalar masses.
This permits systematic exploration over black-hole spin, scalar mass, and coupling structure, which is important for controlling modelling uncertainties in waveform construction and gravitational-wave inference.

\sect{Acknowledgement}This work employs large language models (LLMs), specifically Claude Opus 5.5 for coding, optimization, and documentation, and ChatGPT 5.6 Sol High for manuscript structuring and drafting. The authors maintained strict oversight over all LLM applications and rigorously reviewed all generated content. L.L. was supported by the Beijing Natural Science Foundation (No. IS25014). 
A.K.W.C. thanks Daniela Cros, Gareth Marks, Christopher Moore, Ulrich Sperhake, and Seppe Staelens for helpful discussions.
A.K.W.C. acknowledges full support from the Title-E Fellowship at St John's College, Cambridge, and partial support from the Herchel Smith Fellowship at the University of Cambridge.

\sect{Data availability}The data are available upon reasonable request from the authors.

\bibliography{main}
\onecolumngrid
\appendix*
\setcounter{equation}{0}
\section*{Supplementary Material}
\begin{figure*}[htbp!]
    \centering
    \includegraphics[width=\linewidth]{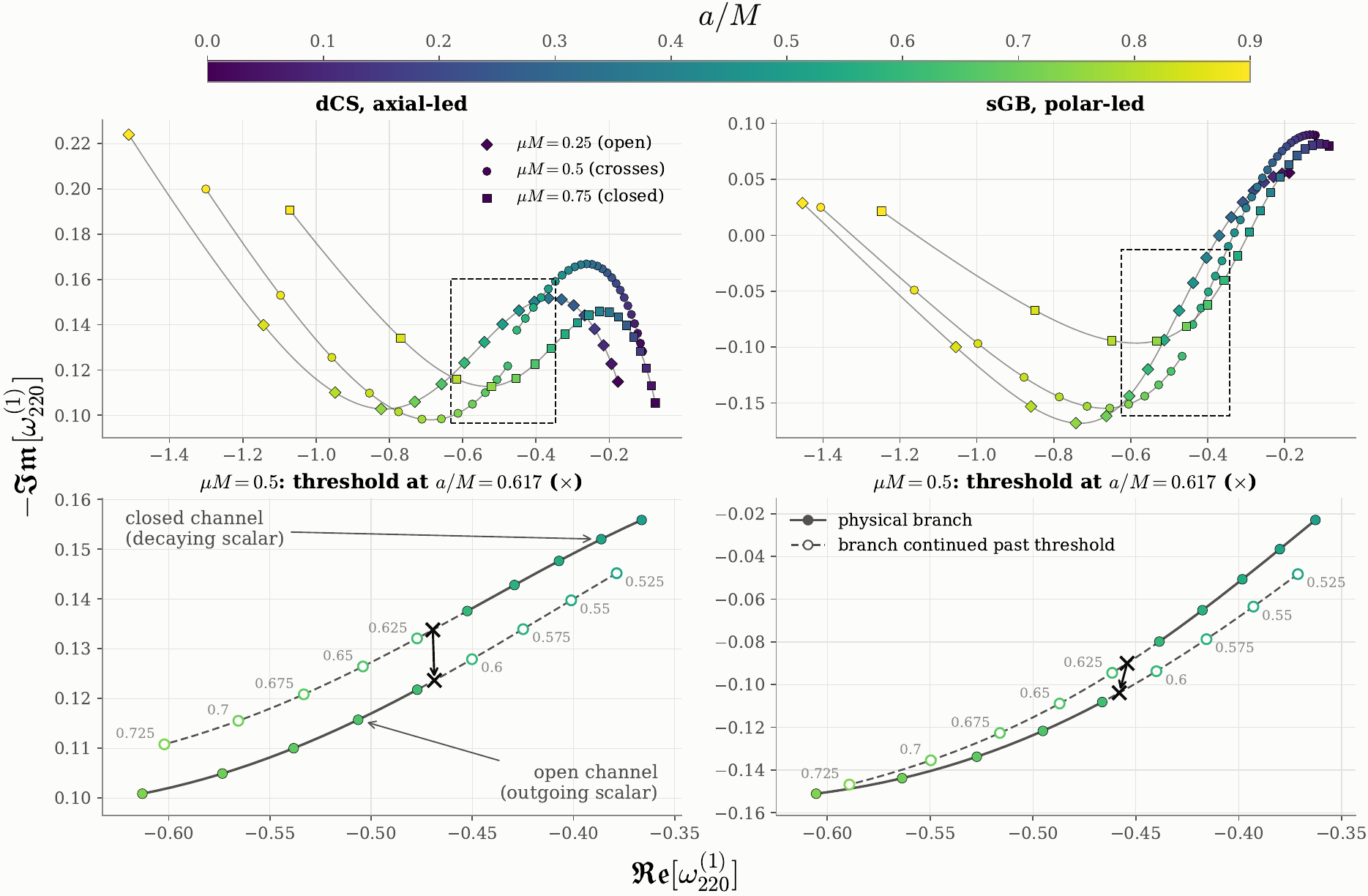}
    \caption{Threshold transition of the $(2,2,0)$ shift in the complex plane, for dynamical Chern-Simons (left; axial-led) and scalar Gauss-Bonnet (right; polar-led) gravity, with the spin as colour. Top: $\mu M=0.25$ (open channel at every spin), $0.5$ (crosses the threshold) and $0.75$ (closed). The box marks the bottom panels. Bottom: the $\mu M=0.5$ crossing. Filled markers are the physical branch, open markers the same branch continued past the threshold, and at $a/M=0.617$ ($\times$) the mode hops between them.}
    \label{fig:threshold}
\end{figure*}
\twocolumngrid

\sect{Channels, the branch rule, and asymptotic factors.}The scalar perturbation satisfies an inhomogeneous massive Klein--Gordon equation throughout the spacetime.
In the asymptotic region, its homogeneous solutions have radial dependence $e^{\pm i k r}$, with $
k=(\omega^2-\mu^2)^{1/2}$,
whereas the gravitational perturbation has asymptotic radial wave number $\omega$.
Because $k(\omega)$ is double valued, with branch points at $\omega=\pm\mu$, specifying the quasinormal-mode boundary condition requires a choice of branch~\cite{Berti_2026}.
We fix this branch by requiring $k(\omega)$ to be analytic in the upper half of the complex-$\omega$ plane, as appropriate for retarded boundary conditions, and analytically continue it to the quasinormal-mode frequency.
For the positive-frequency modes considered here, if $\Re(\omega)>\mu$ (open channel), the continuation gives $\Re(k)>0$ and, for a damped quasinormal mode, $\Im(k)<0$.
This corresponds to an outgoing scalar wave. If $0<\Re(\omega)<\mu$ (closed channel), the continuation instead gives $\Im(k)>0$, so the homogeneous scalar mode is evanescent and decays exponentially at infinity.
In terms of the $\kappa$ defined in the main text,$
k=s\kappa$, where $s=\mathrm{sgn}\!\left[\Re(\omega)-\mu\right]$.
With this continuation convention, the open- and closed-channel continuations are separated in the lower half-plane by a cut extending downward from the branch point $\omega=\mu$~\cite{Leaver_1986,Ching_1995,Burko_2004,De_Amicis_2024,De_Amicis_2025}.

The channel fixes the asymptotic factor of the scalar perturbation. 
In the open channel, it is the free outgoing wave, $\mathcal{P}_s\to e^{ikr}r^{iMc}$ with $c=(2\omega^2-\mu^2)/k$ (Supplemental Material). 
This wave outgrows the metric perturbation, $|\mathcal{P}_s/\mathcal{P}_0|\sim e^{gr}$ with $g=\mathfrak{Im}(\omega^{(0)})-\mathfrak{Im}(k)>0$, increasingly so towards the threshold, so the $\mathcal{O}(\zeta)$ back-reaction of the scalar on the metric is a divergent radial integral. 
Like the norm of a quasinormal mode, it is defined by analytic continuation, and we evaluate it consistently with that continuation (c.f. Supplemental Material). 
In the closed channel, the free scalar decays, and at large $r$ the scalar perturbation is instead the wave forced by the metric perturbation, $\propto e^{i\omega r}$, so it carries the metric factor $\mathcal{P}_0$. 
Factoring out $e^{ikr}$ instead would leave a remainder growing like $e^{(|\mathfrak{Im}(k)|+|\mathfrak{Im}(\omega)|)r}$, which cannot be captured by spectral methods.
With this choice, the closed channel agrees with an independent zero-spin code to $2\times10^{-7}$ at $\mu M=1$.

\sect{Continued branches and finite coupling.}At first order, the channel enters only through the scalar response to the Kerr mode, that is, through the sign of $k$ in its boundary factor, $\omega^{(1)}=\Omega(s\kappa)$ [Eq.~\eqref{eq:threshold}]. Both boundary conditions remain well defined on either side of the threshold, so each branch can be continued past it by imposing the other one. That is, the closed-channel shift $\Omega(-\kappa)$ for $\mathfrak{Re}(\omega^{(0)})>\mu$, and the open-channel shift $\Omega(+\kappa)$ for $\mathfrak{Re}(\omega^{(0)})<\mu$ (open markers in Fig.~\ref{fig:threshold}, hollow markers in Fig.~\ref{fig:omega_1}). Each continued branch extends the corresponding physical branch smoothly through the threshold (Fig.~\ref{fig:threshold}, bottom), as the analyticity of $A$ and $B$ requires, and at $a/M=0.617$ the two branches differ by the jump of Eq.~\eqref{eq:jump}. Whether a mode crosses at all depends on the scalar mass. For the $(2,2,0)$ mode, $\mathfrak{Re}(\omega^{(0)})M$ rises from $0.374$ at $a=0$ to $0.672$ at $a/M=0.9$, so the channel is open at every spin for $\mu M=0.25$, closed at every spin for $\mu M=0.75$, and crosses once for $\mu M=0.5$ (Fig.~\ref{fig:threshold}, top).

At finite $\zeta$, each branch's pole, $\omega_\pm=\omega^{(0)}+\zeta\,\Omega(\pm\kappa)+\mathcal{O}(\zeta^2)$, is physical only on its own side of the cut, so the mode hops where its own real part, rather than that of $\omega^{(0)}$, reaches $\mu$. For $\mu M=0.5$ and $\zeta=0.1$, the closed-channel pole leaves the physical sheet at $a/M\simeq0.772$ and the open-channel pole enters it at $0.776$ (dynamical Chern-Simons, axial-led) and $0.777$ (scalar Gauss-Bonnet, polar-led), instead of both at $0.617$ as $\zeta\to0$ (shown in the narrow window between them) the mode's contribution is carried by the branch-cut integral~\cite{Burko_2004}. The signal itself, $h(t)\propto\int e^{-i\omega t}\tilde{G}(\omega)\,\dd\omega$ along a line above the real axis, on which $\tilde{G}$ is analytic and varies continuously with $a$, stays continuous through the threshold and only its split into pole and branch-cut contributions changes.

\sect{Field equations and ansatz of the background spacetimes.}We work to first order in $\zeta=(\ell/M)^4$, in units with $G=c=M=1$, with the metric signature $(-,+,+,+)$ and in modified Boyer--Lindquist coordinates $(t,r,\chi,\phi)$, where $\chi=\cos\theta$.
The field equations are
\begin{equation}\label{eq:FE}
\begin{split}
& R_\alpha{}^\beta+\zeta \left(\mathscr{A}_\alpha{}^\beta-\bar T_\alpha{}^\beta\right)=0, \\
& \left(\Box-\mu^2\right)\bar{\vartheta}+\mathscr{Q}=0,
\end{split}
\end{equation}
where
\begin{equation}
\begin{split}
[\mathscr{A}_{\rm sGB}]_{\alpha }{}^{\beta} & =
\left(
\delta^{\alpha\sigma\rho\tau}_{\beta\zeta\gamma\delta}
-\frac{1}{2}\delta^{\beta}{}_{\alpha}
\delta^{\eta\sigma\rho\tau}_{\eta\zeta\gamma\delta}
\right)
R^{\gamma\delta}{}_{\rho\tau}
\nabla^{\lambda}\nabla_{\sigma}\bar{\vartheta}, \\
[\mathscr{A}_{\rm dCS}]^{\alpha \beta} & = -4 \left(\nabla_\sigma \nabla_\delta \bar{\vartheta} \right) \tilde{R}^{\delta (\alpha \beta) \sigma},
\end{split}
\end{equation}
and
\begin{equation}
\bar T_\alpha{}^\beta=\frac{1}{2}\nabla_\alpha \bar{\vartheta} \nabla^{\beta} \bar{\vartheta}+\frac{1}{4}\delta_\alpha{}^\beta \mu^2 \bar{\vartheta}^2.
\end{equation}
Strictly speaking, $[\mathscr{A}_{\rm dCS}]^{\alpha\beta}$ also contains terms involving covariant derivatives of the Ricci tensor.
These terms vanish on the Kerr background and, at linear order, when acting on the zeroth-order GR quasinormal mode since $R_{\mu\nu}^{(0)}=\delta R_{\mu\nu}=0$.
They therefore contribute only beyond the leading order in $\zeta$ retained here.

The couplings and normalisation are those of Ref.~\cite{Chung_2026b}. When the coefficients of the source $\mathscr{Q}$, of the curvature coupling and of $\bar T$ are left free and fitted to the dCS data released with that reference, they come out equal to the values in Eq.~\eqref{eq:FE} to within $10^{-10}$, $10^{-8}$ and $10^{-7}$, respectively, and the sGB data pass the same test. The metric is written in the ansatz of Ref.~\cite{Chung_2026b},
\begin{equation}\label{eq:ansatz}
\begin{split}
g_{tt}&=-\left(1-\frac{2r}{\Sigma}-\zeta H_1\right),\\
g_{t\phi} &=-(1+\zeta H_2)\frac{2ar(1-\chi^2)}{\Sigma},\\
g_{rr}&=(1+ \zeta H_3)\frac{\Sigma}{\Delta},\\
g_{\chi\chi} &=(1+ \zeta H_3)\frac{\Sigma}{1-\chi^2},\\
g_{\phi\phi}&=(1+ \zeta H_4)(1-\chi^2)\left(r^2+a^2+\frac{2a^2r(1-\chi^2)}{\Sigma}\right).
\end{split}
\end{equation}

\sect{Numerical representation of the backgrounds}The scalar solves the second of Eqs.~\eqref{eq:FE} on the exact Kerr metric. For $\mu>0$, it does not simply decay like a Yukawa field. Far from the hole, the curvature source is balanced by the mass term, which leaves an algebraic tail $\bar\vartheta\simeq\mathscr{Q}/\mu^2$, equal to $-288a\chi/\mu^2r^7$ in dCS and to $48/\mu^2r^6$ in sGB gravity, on top of an exponentially small Yukawa part. To capture both behaviours, we split the radial domain into panels $r_+=b_0<b_1<\dots<b_K=R$, whose radii roughly double from one panel to the next, with $R=\max(8r_+,32/\mu)$, and we add a compactified exterior in the variable $X=R/r\in(0,1]$,
\begin{equation}
\bar\vartheta=\begin{cases}\sum_{n,l}c^{(i)}_{nl}\,T_n(u_i)\,P_{2l+p}(\chi), & b_i\le r<b_{i+1},\\[2pt] X^{q}\sum_{n,l}e_{nl}\,T_n(2X-1)\,P_{2l+p}(\chi), & r\ge R.\end{cases}
\end{equation}
Here $u_i$ maps the $i$th panel onto $[-1,1]$. The parity index is $p=1$ for the dCS pseudoscalar, which is odd in $\chi$, and $p=0$ for the sGB scalar. The factor $X^q$, with $q=7$ in dCS and $q=6$ in sGB gravity, removes the algebraic tail, so that the series in the exterior is regular at infinity. The metric functions use a single global expansion,
\begin{equation}
H_k=\sum_{n=0}^{N_z}\sum_{l=0}^{N_l}h_{knl}\,T_n(z)\,P_{2l}(\chi),\quad z=\frac{2r_+}{r}-1\in(-1,1].
\end{equation}
All background derivatives up to second order in $r$ and $\chi$ are exact derivatives of these series, which we obtain from the recurrence relations of the Chebyshev and Legendre polynomials, together with the closed-form Kerr terms, which we differentiate symbolically.

The scalar equation is collocated on the Chebyshev--Gauss nodes of each panel and on the positive Gauss--Legendre nodes in $\chi$. The field and its radial derivative are matched at every interface, and the whole system is solved in a single least-squares step. Regularity is built into the basis, since it excludes both the logarithmic branch at the horizon and the growing Yukawa branch at infinity. For the metric, the linearised Ricci operator acting on $H_k$ is obtained exactly by complex-step differentiation of the full Ricci tensor of the ansatz~\eqref{eq:ansatz}. Its eight independent components are collocated on a grid oversampled by a factor of $1.5$, and the overdetermined system is again solved by least squares. The operator has a four-dimensional near-null space, made of an axis mode, rescalings of $t$ and $\phi$, and the Kerr perturbations $\delta M$ and $\delta J$. We remove it with four conditions at infinity. The first two, $H_1(\infty)=0$ and $H_3(\infty)=H_4(\infty)$, fix the normalisation of time and the isotropy of the asymptotic frame. The third, $\lim_{r\to\infty}rH_1=-H_4(\infty)$, keeps the ADM mass unchanged, and the fourth, $H_2(\infty)=-H_4(\infty)/2$, keeps the angular momentum unchanged. Hence $M$ and $J$ are the physical mass and angular momentum of the corrected black hole, and $\omega M$ is expressed in physical units. The production resolutions are $(N_z,N_l)=(40,10)$ for the metric and $(N,N_l,N_x)=(28,10,30)$ for every scalar panel.

For very light fields, a single expansion in $z$ is not adequate. The energy-momentum tensor of the scalar contains the factor $e^{-2\mu r}$, which in terms of $z$ reads $\exp[-4\mu r_+/(z+1)]$. This is an essential singularity at $z=-1$, inside a layer of width $4\mu r_+$ that a Chebyshev series in $z$ resolves only slowly. The ADM-mass condition, which involves the derivative of $H_1$ at $z=-1$, is the most sensitive to it. For $\mu M\lesssim0.02$ the single expansion therefore misplaces the mass normalisation, and in sGB gravity it shifts $\omega^{(1)}$ by up to $1.4\times10^{-4}$. That is negligible for the spectra, but it amounts to $20\%$ of the small difference $\omega^{(1)}(\mu)-\omega^{(1)}(0)$ at $\mu M\le2\times10^{-3}$, and raising $N_z$ does not remove it. For the small-mass runs of Fig.~\ref{fig:small_mu} we therefore expand the metric on three radial domains. The near zone $r_+\le r\le16M$ uses Chebyshev polynomials in $1/r$. The intermediate zone $16M\le r\le40/\mu$ uses Chebyshev polynomials in $\ln r$, in which both the Compton layer at $r\sim1/\mu$ and all power laws are smooth. The far zone $r\ge40/\mu$ uses Chebyshev polynomials in $40/(\mu r)$, since there $e^{-2\mu r}<e^{-80}$ and only the algebraic tail remains. The functions $H_k$ and their radial derivatives are continuous across both interfaces, a condition we impose exactly by expanding the unknown coefficients in the null space of the continuity conditions, and the four conditions at infinity are imposed in the far domain as before. With $32$, $32$ and $16$ radial functions in the three domains, the operator has exactly four near-null directions, separated from the rest of its spectrum by ten orders of magnitude, and increasing the resolution to $40$, $40$ and $20$ changes $\omega^{(1)}$ by less than $10^{-7}$. At $\mu M=0$ and $0.2$, where the single expansion is accurate, the two representations agree to $3\times10^{-7}$.

\sect{Accuracy of the backgrounds}The scalar reproduces the benchmark values of Ref.~\cite{Chung_2026b}. At $a=0.2$ and $\mu=0.1$, for example, we find $\bar\vartheta(r_+,\chi=\tfrac12)=-0.182283541190$, compared with the benchmark value $-0.182283541192$, and relative differences of $3\times10^{-11}$, $10^{-10}$ and $2\times10^{-9}$ at $r=3$, $10$ and $100$. For every background, we also re-derive both field equations from the full Riemann tensor at points that do not belong to the collocation grid. The scalar equation then holds to within $3\times10^{-7}$ of the source envelope, which is $288a/r^7$ in dCS and $48/r^6$ in sGB gravity, everywhere inside $r=30$. The metric equation, measured relative to the size of the source $|\mathscr{A}-\bar T|$, holds to a median of $10^{-8}$ in both theories, with a maximum of $3\times10^{-8}$ for dCS at $a\le0.8$ and of $5\times10^{-8}$ for sGB. At $a=0.9$ the residual of the metric equation is $2$--$5\times10^{-8}$, a limit set by the resolution. The corrections of order $\zeta$ to the ADM mass and angular momentum decay as $1/r^2$ and are below $3\times10^{-7}$ at $r=10^4$. In the massless and slowly rotating limit, at $a=0.005$ and $\mu=10^{-3}$, the code, which solves the full equations without any expansion in the spin, recovers the solution of Ref.~\cite{Yunes_2009} to $10^{-7}$ at the horizon.

\sect{Applicability of the Regge--Wheeler gauge}The scalar mass does not alter the diffeomorphism freedom of the theory.
Under an infinitesimal coordinate transformation generated by $\xi^\mu$,
the perturbations transform as
\begin{equation}
h_{\mu\nu}\rightarrow h_{\mu\nu}
-2\nabla_{(\mu}\xi_{\nu)},\qquad
\delta\vartheta\rightarrow
\delta\vartheta-\xi^\mu\nabla_\mu\bar\vartheta .
\end{equation}
The four components of $\xi^\mu$ may therefore be used to impose the same
Regge--Wheeler-type gauge conditions on the metric perturbation as in the
massless case.  Because the background scalar is nonvanishing, this
transformation generally induces a corresponding change in
$\delta\vartheta$, which remains an independent dynamical variable and is
not gauge-fixed.  Hence the presence of the mass term
$\mu^2\vartheta^2/2$ introduces no additional obstruction to the gauge
choice used here.

\sect{Ansatz of the perturbations}We follow METRICS~\cite{Chung_2023, Chung_2024, Chung_2024b, Lam_2026}, but we build the discrete operator numerically at each collocation point instead of deriving the equations symbolically. In this part, we use the metric signature $(+,-,-,-)$ of METRICS. The metric perturbation is
\begin{equation}\label{eq:pert}
h_{\mu\nu}=\sum_{k=1}^{6}h_k(r,\chi)\,\tau^{(k)}_{\mu\nu}(r,\chi)\,\mathcal{P}_0(r)\,e^{im\phi-i\omega t},
\end{equation}
with
\begin{equation}\label{eq:P0}
\mathcal{P}_0=e^{i\omega r_*}\left(\frac{r-r_+}{r}\right)^{-i(2\omega-m\Omega_H)r_+/b},
\end{equation}
where $b=2\kappa r_+$, and $\kappa$ is the surface gravity of the outer event horizon of the black hole. 
$\mathcal{P}_0$ behaves as the ingoing wave $(r-r_+)^{-i(\omega-m\Omega_H)r_+/b}$ at the horizon and as the outgoing wave $e^{i\omega r}r^{2iM\omega}$ at infinity. 
The template tensors $\tau^{(k)}$ carry fixed powers of $r$, of $(r-r_+)/r$ and of $1-\chi^2$, together with the angular structure. The function $h_1$ appears in $h_{tt}$, $h_2$ in $h_{tr}$ and $h_3$ in $h_{rr}$. A single trace-type function $h_4$ appears in both $h_{\chi\chi}$ and $h_{\phi\phi}$. The functions $h_5$ and $h_6$ are the axial parts of $h_{tA}$ and $h_{rA}$, built from $\partial_\chi$ and $im$ acting on the harmonic, and $h_{\chi\phi}$ vanishes. This is the Regge--Wheeler-type gauge used by METRICS, and it is the only gauge fixing we impose. The scalar perturbation carries its own asymptotic factor,
\begin{equation}\label{eq:Ps}
\begin{split}
&\delta\vartheta=\frac{h_7(r,\chi)}{r}\,\mathcal{P}_s(r)\,e^{im\phi-i\omega t},\quad \mathcal{P}_s\to e^{ikr}\,r^{iMc},\\
&k=\sqrt{\omega^2-\mu^2},\quad c=\frac{2\omega^2-\mu^2}{k},
\end{split}
\end{equation}
with the same ingoing factor at the horizon as $\mathcal{P}_0$. The branch of $k$ is discussed in the section on the massive-scalar asymptotics below. Each unknown is expanded as
\begin{equation}
h_k=\sum_{n=0}^{N_z}\sum_{\ell}v_{kn\ell}\,T_n(z)\,\bar P_\ell^{|m|}(\chi),
\end{equation}
with normalised associated Legendre functions $\bar P_\ell^{|m|}$. The expansion is restricted to one sector of the reflection $\chi\to-\chi$. The polar-led sector contains $h_1,\dots,h_4$ with $\ell+|m|$ even together with $h_5$ and $h_6$ with $\ell+|m|$ odd, and the axial-led sector is the opposite. Being a pseudoscalar, the dCS scalar has the equatorial parity opposite to that of $h_1,\dots,h_4$, whereas the sGB scalar has the same parity. We keep $N_\chi+1$ values of $\ell$ for every function.

\sect{Construction of the discrete operator}At each collocation node the background enters only through its metric jets, that is, through the values of the metric and of its first and second derivatives, which are computed exactly as described above. Every perturbation quantity at the node is a linear form in the radial and angular jets of the unknowns. Its coefficients are polynomials in $\omega$, because $\partial_t\to-i\omega$ and because $\mathcal{P}_0'/\mathcal{P}_0$ is linear in $\omega$. We build three kinds of rows from these forms. The metric rows come from the linearised mixed Ricci tensor $\delta R^{j}{}_{a}$, evaluated for the ten pairs $j\le a$, which yields coefficients $c_{a,k,p,d_r,d_\chi}$ at each node, where $p$ is the power of $\omega$. The couplings of order $\zeta$ come from evaluating $\mathscr{A}^\mu{}_\nu-\bar T^\mu{}_\nu$ and the scalar operator as pointwise functions of the metric, of the scalar and of their first and second derivatives. Their Jacobians with respect to every jet entry are exact to rounding error, since we compute them with complex steps, and contracting them with the perturbation forms gives the parts of the perturbed field equations that involve the metric perturbation and the scalar perturbation. Finally, multiplying the node coefficients by $\dd^{d_r}T_n/\dd r^{d_r}$ and $\dd^{d_\chi}\bar P_\ell/\dd\chi^{d_\chi}$ assembles the rectangular matrix. The linear problem is then $\mathbf{D}(\omega,\zeta)\mathbf{v}=0$ with
\begin{align}\label{eq:D}
\mathbf{D}(\omega,\zeta)&=\mathbf{D}^{(0)}(\omega)+\zeta\,\mathbf{D}^{(1)}(\omega),\nonumber\\
\mathbf{D}^{(0)}&=\begin{pmatrix}\mathbf{E}_0&0\\ \mathbf{Q}_g&\mathbf{Q}_s\end{pmatrix},\quad
\mathbf{D}^{(1)}=\begin{pmatrix}\partial_\zeta\mathbf{E}+\mathbf{S}_g&\mathbf{S}_s\\0&0\end{pmatrix}.
\end{align}
Here $\mathbf{E}_0(\omega)=\sum_{p}\omega^p\mathbf{E}_{0,p}$ is the Kerr operator, with ten rows per node. The blocks $\mathbf{Q}_g$ and $\mathbf{Q}_s$ form the scalar equation, with one row per node, linearised on Kerr in the presence of the background scalar, which the metric perturbation drives through $\delta\mathscr{Q}$ and $\delta\Box$. The block $\partial_\zeta\mathbf{E}$ is the change of the metric operator on the corrected background Kerr$+\zeta H$, and $\mathbf{S}_g$ and $\mathbf{S}_s$ are the dCS or sGB sources. The terms of order $\zeta$ in the scalar equation act on a gravity-led mode only at order $\zeta^2$ and are dropped. We use all ten linearised field equations at every node, without a reduction to master equations and without a separate elimination of the constraints. The Bianchi identities make the overdetermined system consistent, and its least-squares residual saturates at the truncation error.

\sect{Zeroth-order eigenproblem}At $\zeta=0$ the operator $\mathbf{D}^{(0)}$ is block lower triangular, because the Kerr mode drives the scalar but the scalar does not act back. The Kerr frequency and eigenvector solve
\begin{equation}
\mathbf{E}_0(\omega_0)\mathbf{v}_g=0,\qquad v_{g,\,(\mathrm{lead},\,n=0,\,\ell=|m|)}=1.
\end{equation}
The second condition fixes the normalisation by setting to one the zeroth Chebyshev coefficient of the leading function in the $\ell=|m|$ harmonic. The leading function is $h_1$ for polar-led and $h_5$ for axial-led modes. We solve this nonlinear problem by Gauss--Newton iteration on the free coefficients and on $\omega$, with Moore--Penrose least-squares steps, rows scaled to unit maximum, and the stopping rule $|\delta\omega|<10^{-13}|\omega|$~\cite{Chung_2024}. At the production resolution, this reproduces the Kerr $(2,2,0)$ frequencies obtained with Leaver's method, as implemented in Ref.~\cite{Stein_2019}, to a median of $8\times10^{-11}$ and at most $5\times10^{-10}$ over the 37 spins we use. The scalar response to the Kerr mode then follows from the row-scaled least-squares solve $\mathbf{Q}_s\mathbf{v}_s=-\mathbf{Q}_g\mathbf{v}_g$, and the vector $\mathbf{v}=(\mathbf{v}_g,\mathbf{v}_s)$ spans the right null space of $\mathbf{D}^{(0)}(\omega_0)$.

\sect{Boundary conditions, bases, and resolution}There are no explicit boundary rows. The factors $\mathcal{P}_0$ and $\mathcal{P}_s$ contain the ingoing behaviour at the horizon and the outgoing behaviour at infinity, and the Chebyshev expansion of a bounded function on $z\in[-1,1]$ excludes the other branches at both ends. The powers of $(r-r_+)/r$ in the templates make every component regular at the horizon.

Collocation uses the $N_z+1$ Chebyshev--Gauss nodes in $z$ times the $N_\chi+1$ positive nodes of the Gauss--Legendre rule with $2N_\chi+2$ points, and equatorial symmetry covers $\chi<0$. For the $(2,2,0)$ spectrum we use $N_z=24$ in dCS and $N_z=32$ in sGB gravity, with $N_\chi=6$ for $a\le0.5$, $N_\chi=8$ for $a\le0.8$ and $N_\chi=10$ at $a=0.9$. The system has $11(N_z+1)(N_\chi+1)$ rows and $7(N_z+1)(N_\chi+1)$ columns. The $(3,3,0)$ mode uses the same resolutions, and the $(2,2,1)$ overtone uses the same angular resolutions with $N_z=24$ in both theories.

\sect{First-order frequency shift and the perturbation scheme}Writing $\omega=\omega_0+\zeta\omega_1$ and $\mathbf{v}\to\mathbf{v}+\zeta\,\delta\mathbf{v}$ in Eq.~\eqref{eq:D}, the equation at order $\zeta$ is
\begin{equation}\label{eq:first}
\begin{split}
&\mathbf{D}^{(0)}(\omega_0)\,\delta\mathbf{v}+\omega_1\,\partial_\omega\mathbf{D}^{(0)}(\omega_0)\,\mathbf{v}=-\mathbf{D}^{(1)}(\omega_0)\,\mathbf{v},\\
    &\delta v_{(\mathrm{lead},0,|m|)}=0 .\\
\end{split}
\end{equation}
We solve it as a single bordered least-squares problem,
\begin{equation}
\begin{split}
&\mathbf{J}\begin{pmatrix}\delta\mathbf{v}_{\mathrm{free}}\\ \omega_1\end{pmatrix}=-\mathbf{s}\odot\mathbf{D}^{(1)}\mathbf{v},\\
&\mathbf{J}=\mathbf{s}\odot\Big[\mathbf{D}^{(0)}(\omega_0)_{:,\mathrm{free}},\ \partial_\omega\mathbf{D}^{(0)}(\omega_0)\,\mathbf{v}\Big],
\end{split}
\end{equation}
with one fixed row scaling $\mathbf{s}$, taken from $\mathbf{D}^{(0)}(\omega_0)$ and applied at all orders. We do not compute left eigenvectors. The bordering column $\partial_\omega\mathbf{D}^{(0)}\mathbf{v}$ and the normalisation condition give $\mathbf{J}$ full column rank, and as the discretisation converges the solution becomes equivalent to the projection $\omega_1=-\mathbf{u}^\dagger\mathbf{D}^{(1)}\mathbf{v}/\mathbf{u}^\dagger\partial_\omega\mathbf{D}^{(0)}\mathbf{v}$, where $\mathbf{u}$ is the left singular vector of the smallest singular value. Our independent zero-spin code, described below, uses that projection.

The ingredients are computed as follows. The derivative $\partial_\zeta\mathbf{E}$ is obtained from operators $\mathbf{E}_{\pm1}$ and $\mathbf{E}_{\pm2}$ assembled on Kerr$+\zeta H$ at $\zeta=\pm\delta$ and $\pm2\delta$, with $\delta=10^{-6}$, and combined by Richardson extrapolation into $[4(\mathbf{E}_{+1}-\mathbf{E}_{-1})/2\delta-(\mathbf{E}_{+2}-\mathbf{E}_{-2})/4\delta]/3$. The sources $\mathbf{S}_g$ and $\mathbf{S}_s$ follow from the exact Jacobians described above. The derivative $\partial_\omega\mathbf{E}_0$ is analytic, since $\mathbf{E}_0$ is a polynomial in $\omega$. The scalar blocks, in contrast, carry the non-polynomial factor $\mathcal{P}_s/\mathcal{P}_0$ and are built at fixed $\omega$, so their derivative $\partial_\omega\mathbf{Q}$ uses the same Richardson combination with steps $\delta\omega=10^{-4}$ and $2\times10^{-4}$.

\sect{Asymptotic factors at first order}The factor~\eqref{eq:P0} depends on the background through $r_+$, $\kappa$, $\Omega_H$ and $M$, and through the tortoise coordinate. On the corrected background, the horizon stays at $\Delta=0$, because $H_k$ is regular there, whereas $\kappa$ and $\Omega_H$ change at order $\zeta$. We have verified the zeroth law and the rigidity of the corrected horizon. The derivative $\partial_\zeta\Omega_H$ is independent of $\chi$ on the horizon to $10^{-8}$, and $\partial_\zeta\kappa$ to $4\times10^{-8}$ at $a=0.5$ and to $3\times10^{-6}$ at $a=0.8$. Since $M$ and $J$ are unchanged by construction, the factor at infinity is unchanged as well. Each operator in the difference quotients in $\zeta$ is built with its own horizon data, $\kappa(\zeta)$ and $\Omega_H(\zeta)$, and with $r_*'=\sqrt{g_{rr}/N^2}$ evaluated on the axis of the corrected metric, which for Kerr reduces to $(r^2+a^2)/\Delta$. Hence $\partial_\zeta\mathbf{E}$ already contains the first-order variation of both asymptotic factors, the correction $\delta\mathbf{v}$ remains a bounded Chebyshev series at $z=\pm1$, and no separate boundary terms arise. The scalar asymptotic factor $\mathcal{P}_s$ depends only on $\omega$, on $\mu$ and on the Kerr horizon, since the scalar enters at zeroth order.

\sect{Massive-scalar asymptotics, far-zone growth, the threshold, and regularization schemes}The branch rule for $k$, the open and closed scalar channels, and the asymptotic factor used in each are explained in the Supplementary Material of the Letter. Two implementation details are added here. In the closed channel, the scalar carries the metric factor $\mathcal{P}_0$, with the same ingoing behaviour at the horizon, so that its remainder is a bounded Chebyshev series. This choice reproduces the zero-spin code to $2\times10^{-7}$ at $\mu=1$. Because the rule switches exactly at $\mathfrak{Re}\,\omega_0=\mu$, the spectra are computed, fitted and drawn separately on the two sides of the threshold, and the branches continued past the threshold in Figs.~1 and~4 of the Letter are obtained by imposing the asymptotic factor of the other channel at the same spin and mass.

In the open channel, the scalar outgrows the metric wave, $|\mathcal{P}_s/\mathcal{P}_0|\sim e^{gr}$ with $g=\mathfrak{Im}\,\omega_0-\mathfrak{Im}\,k>0$. The back-coupling $\mathbf{S}_s\mathbf{v}_s$ therefore turns the first-order shift into a radial integral that diverges on the real $r$ axis. It is defined by analytic continuation, the same continuation that defines the norm of a quasinormal mode, and we evaluate it with one of three methods, depending on how fast the scalar grows.

Far from the threshold, where $g\cdot300<10$, a condition that covers $\mu\le0.2$ and every closed-channel point, we multiply the back-coupling rows of order $\zeta$ by the window $W=e^{-(r/r_c)^8}$ with $r_c=300$. The result changes by less than $10^{-6}$ when $r_c$ is varied between $200$ and $400$, and it agrees with the zero-spin code to between $10^{-8}$ and $3\times10^{-6}$ for $\mu\le0.2$.

Closer to the threshold, this window has only an optimal-truncation plateau. On a dense one-dimensional grid its accuracy is about $10^{-3}$ at $\mu/\mathfrak{Re}\,\omega_0\simeq0.94$, and beyond that it diverges like $e^{gr_c}$. There we instead use the Zel'dovich factor~\cite{Zeldovich_1961}, $W=e^{-(r/R)^2}$. It is analytic and decays for $|\arg r|<45^\circ$, where the radial contour can be rotated until every outgoing wave decays. The windowed $\omega_1(R)$ therefore equals its contour-rotated value and tends to the analytically continued $\omega_1$ as $R\to\infty$, with a regular expansion in $1/R^2$. We compute $\omega_1(R)$ for $R=5,\dots,16$,  a quartic polynomial in $1/R^2$ to the values with $R=5$--$12$, and take as the error estimate the spread between this fit, a quartic fit to $R=6$--$16$, and a cubic fit. Small widths keep the windowed source negligible at the few outermost radial nodes. In the zero-spin code, this procedure reproduces the cutoff-independent value at $\mu=0.2$ to $3\times10^{-6}$, and it reproduces the extrapolation to large widths, up to $R=35$, to within $10^{-4}$ up to $\mu/\mathfrak{Re}\,\omega_0=0.94$ and to $2$--$5\times10^{-4}$ at $0.99$. The two-dimensional code passes the same tests. At $a=0$ its extrapolated values agree with the zero-spin code to $4\times10^{-5}$ at $\mu=0.33$, where $\mu/\mathfrak{Re}\,\omega_0=0.88$, and to about $10^{-4}$ at $\mu=0.35$, where the ratio is $0.94$. At $\mu=0.37$, where it is $0.99$, they agree within the uncertainty of the zero-spin code itself, which is about $5\times10^{-4}$. Far from the threshold, at $a=0.5$ and $\mu=0.2$, they reproduce the value obtained with the $r^8$ window to $4\times10^{-6}$. Changing $N_z$ from $24$ to $32$ moves them by at most $3\times10^{-4}$, and the fit spreads are at most $6\times10^{-5}$. For a spinning hole just below the threshold, at $a=0.2$ and $\mu=0.4$, where $\mu/\mathfrak{Re}\,\omega_0=0.995$, the extrapolation agrees to $3\times10^{-5}$ with the plateau of the $r^8$ window at $r_c$ between $50$ and $150$. Larger values of $r_c$ give results that are wrong by up to $0.12$, because they switch on the outermost radial nodes, where the window resolves the growing far zone with only one or two points. All spectrum points with $g\cdot300\ge10$, which number $64$ per theory and lie at $\mu=0.3$--$0.6$, therefore use the Zel'dovich extrapolation.

The third method is exterior complex scaling, which we use where the window converges too slowly. The scalar perturbation is then given a radial grid of its own. On the real interior, up to a radius $R_0$, it is expanded in Chebyshev polynomials of $\ln r$ with $N_i=48$ nodes. Beyond $R_0$, the radial coordinate is continued along the line $r=R_0+\rho\,e^{i\theta}$, on which the scalar obeys the free massive Klein--Gordon equation in closed form and the outgoing wave decays. The rotation angle is $\theta=0.4-\arg k$, restricted to $0.9\le\theta\le1.45$, and the frequency shift is obtained from the metric rows alone. The need for this method comes from the part of the scalar that is forced by the metric perturbation. In the open channel it behaves as $e^{i(\omega-k)r}$ relative to $\mathcal{P}_s$, an oscillation that a Chebyshev series on the windowed real axis represents poorly, so that with the $r^8$ window the shift converges only algebraically in $N_z$, by about one decade per ten radial orders. The effect is largest in the parity that couples to the scalar at zero spin, which for the $(2,2,0)$ mode in sGB gravity is the polar-led one. For these modes, at $\mu<0.75\,\mathfrak{Re}\,\omega_0$, the window values differ from the complex-scaled ones by a median of $1.4\times10^{-4}$ and at most $9\times10^{-4}$, while two independent sets of complex-scaled values agree to $2\times10^{-6}$. All $86$ polar-led sGB points of the $(2,2,0)$ spectrum in this range therefore use complex scaling, with $R_0=20/\mu$ restricted to $250\le R_0\le2\times10^4$ and $N_z=32$. For dCS modes of both parities and for axial-led sGB modes, the two methods agree to $10^{-6}$--$10^{-5}$, and the window values are kept. Very close to the threshold, complex scaling converges slowly as well, with a plateau of $2$--$3\times10^{-5}$ at $a=0.5$ and $\mu=0.4$, and there the Zel'dovich extrapolation is retained. Complex scaling is also used for the open-channel points of the $(2,2,1)$ overtone and for the small-mass runs, as described below.

\sect{Convergence and numerical uncertainty}We examine the convergence of $\omega^{(1)}_{220}$ in the radial order $N_z$, which is the resolution that limits the accuracy, at the spins $a/M=0.1$, $0.5$, $0.7$, $0.8$ and $0.9$, in both parities, and at the scalar masses $\mu M=0.2$, $0.5$, $0.6$ and $0.9$, together with $\mu M=0.01$ in dCS and $\mu=0$ in sGB gravity. These masses sample the open channel on both sides of the threshold, which for the $(2,2,0)$-mode shifts from $\mu M=0.37$ at $a=0$ to $0.67$ at $a/M=0.9$, and the closed channel. For each case, we compute $\omega^{(1)}$ at $N_z=10$, $12$, $14$, $16$, $20$, $24$, $28$, $32$ and $36$ on accurate backgrounds, with $(N_z,N_l)=(48,14)$ for the metric and $(36,14,40)$ for the scalar, so that the background does not limit the test. Points near the threshold use the Zel'dovich extrapolation at every order, and polar-led sGB points in the open channel below $0.75\,\mathfrak{Re}\,\omega_0$ use complex scaling, exactly as their production values do.
\begin{figure*}[t]
    \centering
    \includegraphics[width=\linewidth]{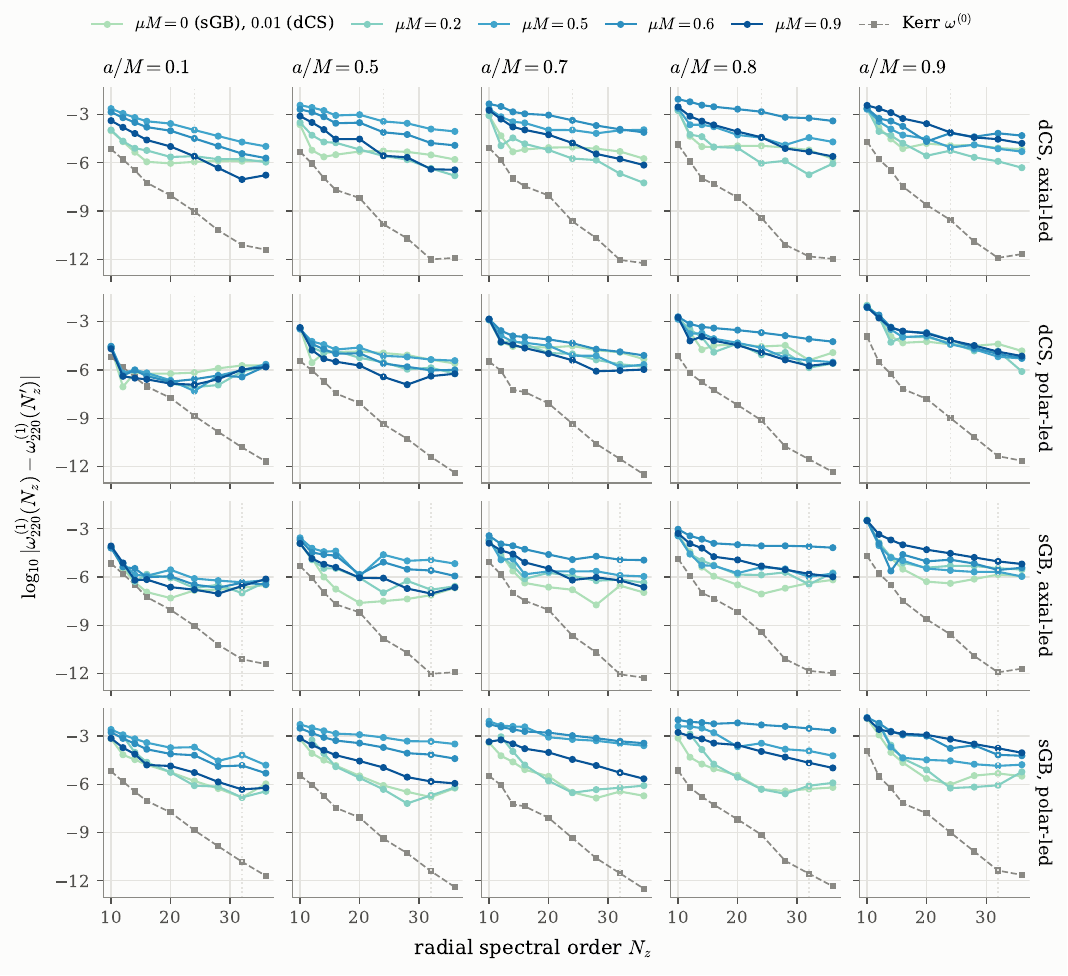}
    \caption{Radial convergence of the $(2,2,0)$ shift. Each panel shows the base-10 logarithm of the backward difference $\mathcal{B}(N_z)$ between successive radial orders for the scalar masses in the legend. The rows are the axial-led and polar-led modes in dCS and in sGB gravity, and the columns are the spins. The grey squares show the same quantity for the Kerr frequency $\omega^{(0)}$, and the dotted vertical line marks the production order.}
    \label{fig:conv}
\end{figure*}
\begin{figure*}[t]
    \centering
    \includegraphics[width=\linewidth]{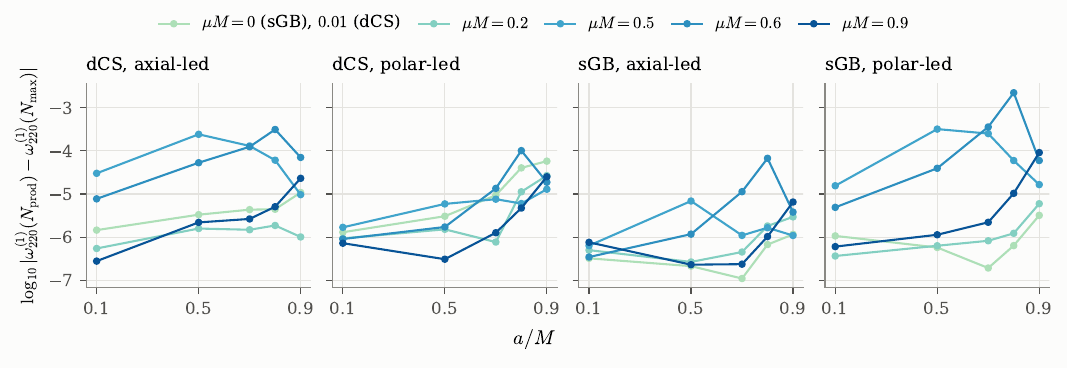}
    \caption{Numerical uncertainty of the production values, $|\omega^{(1)}_{220}(N_{\rm prod})-\omega^{(1)}_{220}(N_{\max})|$ with $N_{\max}=36$, against spin for the scalar masses of Fig.~\ref{fig:conv}, for each theory and parity. The production order is $N_{\rm prod}=24$ in dCS and $32$ in sGB gravity.}
    \label{fig:uncertainty}
\end{figure*}
\begin{figure*}[t]
    \centering
    \includegraphics[width=\linewidth]{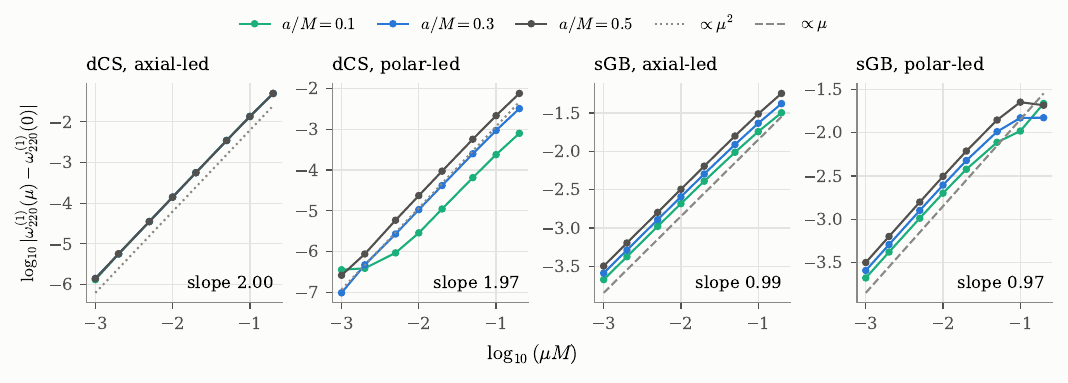}
    \caption{Approach to the massless limit of the $(2,2,0)$ shift at $a/M=0.1$, $0.3$ and $0.5$, for the axial-led and polar-led modes in dCS and in sGB gravity. The grey guide lines have slopes of two in dCS and one in sGB gravity, and each panel gives the mean slope fitted over $0.002\le\mu M\le0.02$.}
    \label{fig:small_mu}
\end{figure*}

Figure~\ref{fig:conv} shows the backward modulus differences $\mathcal{B}(N_z)=|\omega^{(1)}(N_z)-\omega^{(1)}(N_z')|$, where $N_z'$ is the preceding order of the list. The Kerr frequency converges exponentially and reaches the double-precision floor of about $10^{-12}$ near $N_z=32$. The shifts converge more slowly and level off between $10^{-7}$ and $10^{-6}$. That floor is set by the background rather than by the perturbation. The first-order system is overdetermined, and the small residual of the background solved, of $\mathcal{O}(10^{-8})$, moves $\omega^{(1)}$ by about $10^{-6}$. 
A background of higher resolution does not lower the floor further, because the background solved is itself at its double-precision limit. Convergence is slowest in the parity that couples to the scalar at zero spin and near the threshold, for example at $\mu M=0.5$ and $0.6$ for $a/M\ge0.5$, where the far zone of the scalar is the most difficult to be approximated, and it is slowest of all for the polar-led sGB mode at $a/M=0.8$ and $\mu M=0.6$, which lies in the closed channel just above the threshold.

We quote as the numerical uncertainty of a production value the difference $|\omega^{(1)}(N_{\rm prod})-\omega^{(1)}(N_{\max})|$ between the production order, $N_{\rm prod}=24$ in dCS and $32$ in sGB gravity. The highest spectral order we explored is $N_{\max}=36$. It is a direct estimate of the error of the number that enters our spectra and fits, and Fig.~\ref{fig:uncertainty} shows it for all $100$ series of Fig.~\ref{fig:conv}. Its median is $3\times10^{-6}$, it is below $10^{-5}$ in $68\%$ of the cases, and it reaches its maximum, $2.2\times10^{-3}$, for the polar-led sGB mode at $a/M=0.8$ and $\mu M=0.6$. The largest values all lie next to the threshold at high spin. Another possible measure is the smallest backward difference over all computed orders, $\min\mathcal{B}$, which indicates the level at which a series saturates. Its median is $1.2\times10^{-6}$, about half of our estimate. We do not use it as the error of the production values, for two reasons. Once a series has reached its floor, $\mathcal{B}$ fluctuates from order to order, so that its minimum is biased low. Moreover, the minimum often occurs between two orders other than $N_{\rm prod}$, and then it does not describe the value we actually use. Near the threshold, where some series are still decreasing at $N_z=36$, both measures underestimate the true error, and the values there should be regarded as accurate only to the order of the quoted uncertainty.

Several further checks support these estimates. An independent Newton-Raphson solver of the truncated system at finite coupling, $\zeta=\pm10^{-3}$ and $\pm2\times10^{-3}$, gives the slope $[\omega(\zeta)-\omega(-\zeta)]/2\zeta$, which agrees with $\omega^{(1)}$ to $10^{-5}$ after interpolation to $\zeta\to0$. Including the algebraic tail of the scalar and using our own metric correction changes $\omega^{(1)}$ by at most $2.5\times10^{-6}$. Replacing our metric by that of Ref.~\cite{Chung_2026b} changes it by $7\times10^{-6}$ at $a=0.5$ and $\mu=0.1$ and by $1.3\times10^{-4}$ for the polar-led mode at $a=0.8$ and $\mu=0.2$, comparable to the resolution error there. We also rederived the axial $\ell=2$ system on the Schwarzschild background from Eq.~\eqref{eq:FE}, whose vacuum parts are exactly the Regge--Wheeler and massive Klein--Gordon equations, and solved it with a separate one-dimensional code. The full two-dimensional code at $a=0$ agrees with it to $10^{-8}$ for $\mu\le0.01$, to $10^{-7}$ at $\mu=0.1$, to $3\times10^{-6}$ at $\mu=0.2$, and to $2\times10^{-7}$ in the closed channel at $\mu=1$. Finally, at the spins $a=0.3$--$0.753$ used in Ref.~\cite{Chung_2025}, increasing the angular resolution from $N_\chi=8$ to $12$ and the angular order of the background from $N_l=10$ to $14$ shifts the massless dCS $\omega^{(1)}$ by at most $6\times10^{-7}$, and increasing the radial order shifts them by at most $4\times10^{-7}$.

\sect{Approach to the massless limit}Figure~\ref{fig:small_mu} extends Fig.~2 of the Letter to the spins $a/M=0.1$, $0.3$ and $0.5$. All points with $0<\mu M\le0.1$ and the massless references were computed on the three-domain background described above and with the complex-scaled scalar, at $N_z=24$, $N_i=48$ and $R_0=20/\mu$ restricted to $250\le R_0\le2\times10^4$. In dCS gravity the difference $\omega^{(1)}(\mu)-\omega^{(1)}(0)$ grows as $\mu^2$, and in sGB gravity its magnitude grows linearly in $\mu$, in both parities. The only exception is the polar-led dCS mode at $a/M=0.1$, whose shift is so small that below $\mu M\approx5\times10^{-3}$ the difference falls to the numerical floor of about $3\times10^{-7}$ and the curve flattens.

The linear term in sGB gravity can be estimated as follows. Far from the hole, at $r\gg M$, the background scalar is the screened monopole $\bar\vartheta=Q\,e^{-\mu r}/r$, where $Q$ is the scalar charge of the massless scalar field. Our solver gives $Q=4\sqrt{1-a^2}/\big(1+\sqrt{1-a^2}\big)$ to $10^{-5}$, which is 8 times the surface gravity of the Kerr horizon in units of $1/M$. Closer to the hole, for $r\ll1/\mu$, the massive field differs from the massless one by the constant $-Q\mu$ at first order in $\mu$. A constant shift of $\bar\vartheta$ does not affect the near-zone geometry, because the Gauss--Bonnet invariant is a total derivative in four dimensions and $\bar T$ depends on $\bar\vartheta$ only through its gradient and through $\mu^2\bar\vartheta^2$, which is of second order. The only effect at first order in $\mu$ is therefore the change of the field energy stored far from the hole. With the normalisation of Eq.~\eqref{eq:FE}, the Einstein equations read $G_{\mu\nu}=\tfrac{\zeta}{2}\,T_{\mu\nu}$ for the canonical energy-momentum tensor of $\bar\vartheta$, so the mass function obeys $m'=\zeta r^2\rho/4$ with the energy density $\rho=\tfrac12\big[(\partial_r\bar\vartheta)^2+\mu^2\bar\vartheta^2\big]$. For the screened monopole, the gradient and mass terms combine into
\begin{equation}
\int_{r_0}^{\infty}r^2\rho\,\dd r-\left[\int_{r_0}^{\infty}r^2\rho\,\dd r\right]_{\mu=0}=-\frac{Q^2\mu}{2}+\mathcal{O}(\mu^2,\mu^2r_0),
\end{equation}
in which the logarithmic contributions of the two terms cancel. The field thus stores the energy $\zeta Q^2\mu/8$ less than in the massless case. At fixed ADM mass, this energy is absorbed by the black hole, whose mass rises by $\zeta Q^2\mu/8$, while its angular momentum is unchanged, because a static axisymmetric scalar carries no angular momentum. The quasinormal-mode frequency of the hole, $\omega^{(0)}(a_h)/M_h$ with $a_h=J/M_h^2$, then shifts by
\begin{equation}\label{eq:sgb_exact_sm}
\omega^{(1)}(\mu)-\omega^{(1)}(0)=-\frac{Q^2}{8}\left(\omega^{(0)}+2a\,\frac{\partial\omega^{(0)}}{\partial a}\right)\mu+\mathcal{O}(\mu^2),
\end{equation}
which is Eq.~\eqref{eq:sgb_small_mu} of the Letter. The ratio of the computed difference to this prediction is consistent to within $1\%$ at every spin and tends to one linearly in $\mu$, as expected from the next order. At $\mu M=10^{-3}$ it differs from one by $0.02\%$ to $0.5\%$ for $a/M=0.1$, $0.3$ and $0.5$ in both parities. In dCS gravity, the hair is a dipole, $\bar\vartheta\propto r^{-2}$, whose energy far from the hole changes only at order $\mu^3$, and the leading correction is the ordinary term of order $\mu^2$ that the mass introduces into the scalar equation and into the perturbation equations.

\begin{figure}[t]
    \centering
    \includegraphics[width=\linewidth]{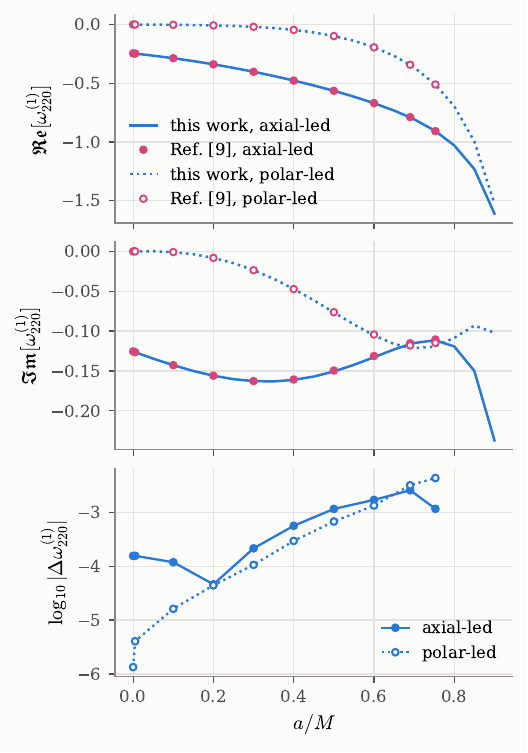}
    \caption{The $(2,2,0)$ shift in dCS gravity at $\mu=0$ from this work, shown as lines, and from Table~II of Ref.~\cite{Chung_2025}, shown as markers, for axial-led and polar-led modes. The top and middle panels show the real and imaginary parts, and the bottom panel shows the base-10 logarithm of the modulus of the difference $\Delta\omega^{(1)}_{220}$ between the two.}
    \label{fig:cly}
\end{figure}

\sect{Consistency checks via comparison with Chung, Lam and Yunes}In the massless limit, our dCS shifts can be compared with the values computed by Chung, Lam and Yunes~\cite{Chung_2025} with the original implementation of METRICS. We compare with the computed values listed in their Table~II, because a fit adds an error of its own. Figure~\ref{fig:cly} shows both calculations and their difference. The axial-led shifts differ by $1.6\times10^{-4}$ at $a=0$ and by $4.6\times10^{-5}$ at $a=0.2$, and from there the difference grows with spin, to $2.6\times10^{-3}$ at $a=0.69$ and $1.2\times10^{-3}$ at $a=0.753$. The polar-led shifts, which vanish at $a=0$, differ by $4\times10^{-6}$ at $a=0.005$, and the difference grows to $4.4\times10^{-3}$ at $a=0.753$. Relative to the size of the shift, the differences are at most $0.3\%$ for axial-led and $0.9\%$ for polar-led modes, so the two calculations agree to better than one percent over the whole range.

The differences are larger than our numerical uncertainty, which at these spins is below $10^{-6}$ in both the radial and the angular resolution, and at most spins they are also larger than the uncertainties listed in Table~II of Ref.~\cite{Chung_2025}. At $a=0$, their origin is clear. Reference~\cite{Chung_2025} extracts the shift from complete solutions at small but finite coupling, as $[\omega(\zeta)-\omega(0)]/\zeta$, which also contains the second-order term $\zeta\,\omega^{(2)}$. Repeating this extraction with our zero-spin code, which can also solve the coupled equations at finite $\zeta$, gives $[\omega(10^{-4})-\omega(0)]/10^{-4}=-0.2460286-0.1254654\,i$. This lies within $2\times10^{-5}$ of their tabulated value, whereas the exact first-order value is $-0.2461427-0.1255990\,i$, and the difference corresponds to $\omega^{(2)}=1.14+1.34\,i$. At finite spin, the second-order term cannot be computed consistently in either code, because it requires the background at second order in $\zeta$. A one-sided extraction in our code is, however, sensitive to it at the level of the differences in Fig.~\ref{fig:cly}, whereas the symmetric slope $[\omega(\zeta)-\omega(-\zeta)]/2\zeta$, which cancels it, reproduces our first-order values to $10^{-6}$ for $a\le0.5$ and to $3\times10^{-5}$ at $a=0.753$. Together with the truncation of the spin expansion of the background in Ref.~\cite{Chung_2025}, this is the most likely origin of the differences at finite spin.

\begin{figure*}[t]
    \centering
    \includegraphics[width=\linewidth]{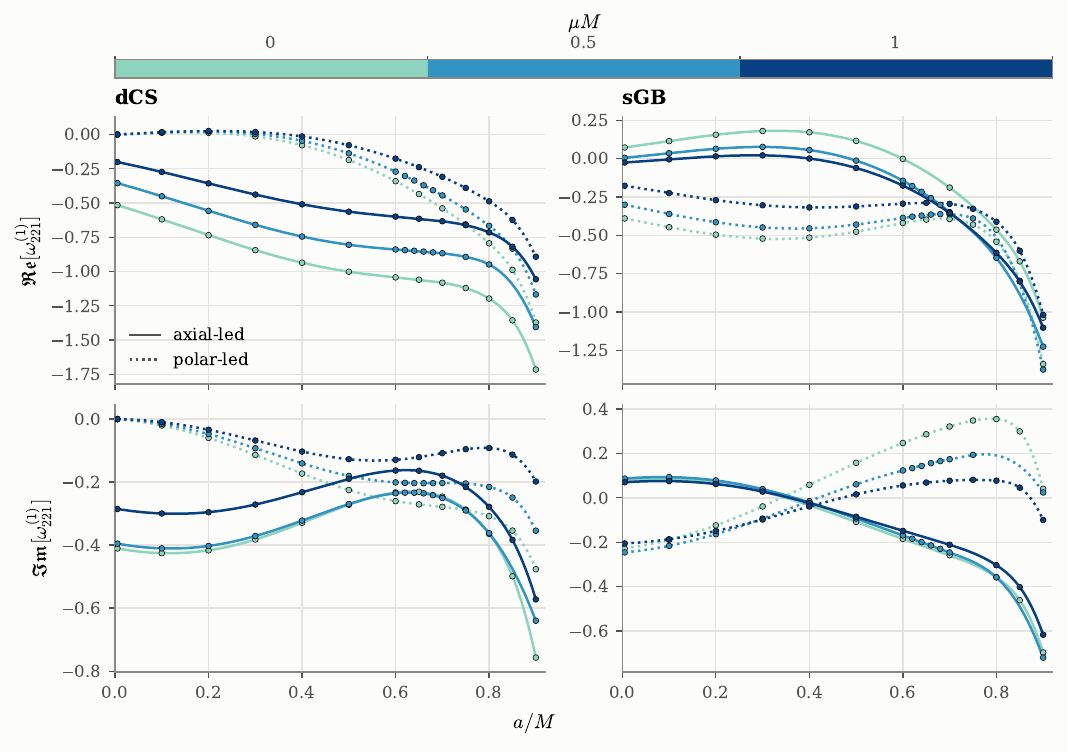}
    \caption{Same as Fig.~\ref{fig:omega_1} of the Letter, at the same scalar masses, but for the $(2,2,1)$ overtone. Open-channel points use the complex-scaled scalar, and points that did not pass the convergence test near the threshold are omitted.}
    \label{fig:omega_1_221}
\end{figure*}

\begin{figure*}[t]
    \centering
    \includegraphics[width=\linewidth]{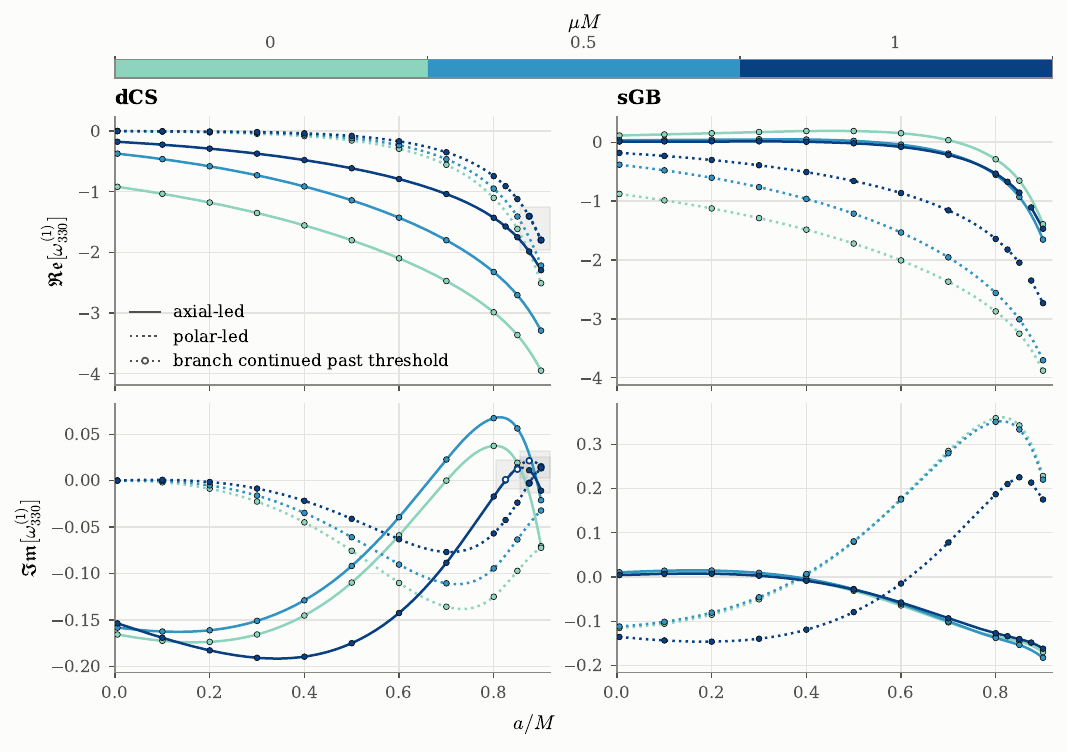}
    \caption{Same as Fig.~\ref{fig:omega_1} of the Letter, at the same scalar masses, but for the $(3,3,0)$ mode.}
    \label{fig:omega_1_330}
\end{figure*}

\sect{Spectra of the $(2,2,1)$ and $(3,3,0)$ modes}Figures~\ref{fig:omega_1_221} and~\ref{fig:omega_1_330} show $\omega^{(1)}$ of the $(2,2,1)$ and $(3,3,0)$ modes in the same format as Fig.~\ref{fig:omega_1} of the Letter. The $(3,3,0)$ mode is computed with the same methods and resolutions as the $(2,2,0)$ mode, at $13$ spins between $a=0.005$ and $0.9$ and at $\mu M=0$, $0.25$, $0.5$, $0.75$ and $1$, with additional spins around each threshold crossing. Its threshold frequency $\mathfrak{Re}\,\omega^{(0)}_{330}$ rises from $0.599$ at $a=0$ to $1.045$ at $a/M=0.9$, so the line $\mu M=0.75$ crosses the threshold at $a/M=0.52$ and the line $\mu M=1$ crosses it at $a/M=0.87$. The branches that continue past these crossings are shown as hollow markers, and open-channel points close to the threshold use the Zel'dovich extrapolation. A pilot study in dCS gravity at $a/M=0.5$ and $\mu M=0.01$ shows convergence to $5\times10^{-5}$ over $N_z=16$--$28$ and agreement with the massless results of Ref.~\cite{Chung_2025} to $10^{-3}$. The $(3,3,0)$ shifts follow the pattern of the $(2,2,0)$ mode but are more significant. In dCS gravity, the real part of the axial-led shift reaches $-4$ at $a/M=0.9$ in the massless limit, and the scalar mass reduces it to about $-2.3$ at $\mu M=1$. In sGB gravity, the polar-led shift dominates, with a real part that also reaches $-4$ at high spin and an imaginary part that changes sign near $a/M=0.4$, while the axial-led shift stays below $0.2$ in modulus up to $a/M=0.7$.

The overtone requires more care. Its scalar response is not captured by the windowed real axis in the open channel, where the window method lets an ingoing scalar solution leak into the result, and sGB shifts computed that way did not converge in $N_z$ at all. All open-channel points of the $(2,2,1)$ spectrum therefore use complex scaling. Because the overtone is strongly damped, the part of the scalar that is forced by the metric perturbation decays relative to the outgoing factor only as $e^{-gr}$, with a low rate $g=\mathfrak{Im}\,(\omega_0-k)$. The exterior must therefore start far out, at $R_0=8/g$ restricted to $250\le R_0\le1500$. Near the threshold, for $\mu\ge0.75\,\mathfrak{Re}\,\omega_0$, a large $R_0$ requires a finely resolved interior, and we use $R_0=10/g$ restricted to $80\le R_0\le160$ with $N_i=56$ nodes; we repeat the computation at $1.25\, R_0$, and accept the point only if both least-squares residuals are below $10^{-6}$, and the two values agree to $1\%$. Points that fail this test are left out of the figure. In the closed channel, the scalar decays, and the window method is used. With complex scaling the sGB overtone converges to $5\times10^{-4}$, and two exterior radii, $R_0=160$ and $240$, agree to $7\times10^{-5}$. The overtone shifts are about twice as large as those of the fundamental mode at low spin. In sGB gravity, the imaginary part of the polar-led shift changes sign near $a/M=0.3$ and reaches $+0.35$ near $a/M=0.8$ for light fields, so that in this parity the scalar coupling makes the overtone longer-lived at high spin.

\end{document}